\documentclass{amsart}
\usepackage{a4wide}
\usepackage{amsfonts}
\usepackage{amsmath}
\usepackage{amsthm}
\usepackage{graphicx}

\newcommand{\Mk}{\ensuremath{{\mathcal M}}}
\newcommand{\Mp}{\ensuremath{{\mathcal M}_p}}
\newcommand{\Mr}{\ensuremath{{\mathcal M}_r}}

\newcommand{\Pn}{\ensuremath{{\mathcal P}_n}}

\newcommand{\Rn}{\ensuremath{{\mathcal R}_n}}

\newcommand{\Tn}{\ensuremath{{\mathcal T}_n}}

\newcommand{\nc}{x}

\newcommand{\FF}{{\bf F}}
\newcommand{\E}{{\bf E}}
\newcommand{\XX}{{\bf X}}

\newcommand{\II}{{\bf I}}
\newcommand{\mm}{{\bf m}}
\newcommand{\yy}{{\bf y}}
\newcommand{\uu}{{\bf u}}
\newcommand{\aaa}{{\bf a}}
\newcommand{\bbb}{{\bf b}}
\newcommand{\dd}{{\bf d}}

\newcommand{\Ord}[1]{\ensuremath{{\mathcal O}\left(#1\right)}}
\newcommand{\mmu}{\mbox{\boldmath{$\mu$}}}

\newtheorem{theorem}{Theorem}
\newtheorem{lemma}{Lemma}
\newtheorem{proposition}{Proposition}

\theoremstyle{definition}

\title{The Distribution of Patterns in Random Trees}
\author{Fr\'ed\'eric Chyzak$^*$, Michael Drmota$^{**}$, Thomas
Klausner$^{**}$, and Gerard Kok$^{***}$}
\date{\today}
\thanks{${}^*$\,INRIA-Rocquencourt, F-78153 Le Chesnay cedex, France, e-mail:
\mbox{frederic.chyzak@inria.fr}}
\thanks{${}^{**}$\,Institut f\"ur Diskrete Mathematik und Geometrie, Technische
Universit\"at Wien,
Wiedner Hauptstra{\ss}e 8-10/113, A-1040 Wien, Austria, e-mail:
\mbox{michael.drmota@tuwien.ac.at},
\mbox{klausner@dmg.tuwien.ac.at}}
\thanks{${}^{***}$\,Delft Institute of Applied Mathematics, Delft University of
Technology,
Mekelweg 4, NL-2628 CD Delft, The Netherlands, e-mail:
\mbox{gkok@fsmat.at}}
\thanks{This research was supported by the Austrian Science Foundation
FWF, grants S8302 and S9604, and by the European {\sc Amadeus} project.}

\begin{document}

\begin{abstract}
Let \Tn{} denote the set of unrooted labeled trees of size $n$ and let
\Mk{} be a particular (finite, unlabeled) tree.
Assuming that every tree of \Tn{} is equally likely, it is shown that
the limiting distribution as $n$~goes to infinity of the number of
occurrences of \Mk{} as an induced subtree is asymptotically normal
with mean value and variance asymptotically equivalent to $\mu n$
and~$\sigma^2n$, respectively, where the constants $\mu>0$
and~$\sigma\ge 0$ are computable.
\end{abstract}

\maketitle

\section{Introduction}
In this paper we consider unrooted labeled trees and analyse the
number of occurrences of a tree pattern as an induced subtree of a
random tree.  It is well known that a typical tree in \Tn, the set of
unrooted labeled trees of size $n$, has about $\mu_k n$ nodes of
degree $k$, where $\mu_k = 1/e(k-1)!$. Moreover, for any fixed $k$ the
total number of nodes of degree $k$ over all trees in \Tn{} satisfies
a central limit theorem with mean and variance asymptotically
equivalent to $\mu_k n$ and~$\sigma_k^2 n$ (for a specific
constant $\sigma_k>0$).  See~\cite{DrGi99}, where Drmota and
Gittenberger explored this phenomenon for unrooted labeled trees and
other types of trees.

A node of degree~$k$ is an occurrence of what can be called a star
with $k$~edges.  In this paper we continue this idea. We consider a
pattern \Mk{}, a given finite tree, and compute the limiting
distribution of the number of occurrences of \Mk{} in \Tn{} as
$n\to\infty$. Note also that there can be overlaps of two or more
copies of \Mk, which we intend to count as separate occurrences.

Our main result in this paper is:
\begin{theorem}\label{Th1}
Let \Mk{} be a given finite tree.  Then the limiting distribution of
the number of occurrences of \Mk{} (as induced subtrees) in a tree of
\Tn{} is asymptotically normal with mean and variance asymptotically
equivalent to $\mu n$ and~$\sigma^2 n$, respectively, where
$\mu>0$ and $\sigma^2\ge0$ depend on the pattern
\Mk{} and can be computed explicitly and 
algorithmically and can be represented as polynomials (with rational
coefficients) in $1/e$.
\end{theorem}

We consider here a random variable $X$ as Gaussian if its
characteristic function is given by ${\bf E}\, e^{itX} = e^{i\mu t-
\sigma^2 t^2/2}$, that is, the case of zero variance $\sigma^2 = 0$ is
included here.  For example, if \Mk{}~consists just of one edge (and
two nodes), then the number of occurrences of \Mk{} in \Tn{} is $n-1$
and thus constant.  So in that particular case we have $\mu = 1$ and
$\sigma^2 = 0$. Nevertheless we conjecture that $\sigma^2 >0$ in
all other cases.

\medskip

As already mentioned, the case of stars (or nodes of given degree)
has been discussed in \cite{DrGi99} for various classes of trees.
Some previous work for unlabeled trees is due to Robinson and
Schwenk \cite{RS75}. Patterns in (rooted planar) trees have also
been considered by Dershowitz and Zaks \cite{MR91a:05033} under the
limitation that patterns start at the root.  In a work on patterns
in random binary search trees, Flajolet, Gourdon, and Mart\'\i nez
\cite{FlGoMa97} obtained a central limit theorem.  Flajolet and
Steyaert also analysed an algorithm for pattern matchings in trees
\cite{MR82d:68023,MR82g:68030,MR86h:68070}. Further Ruci{\`n}ski
\cite{Ru88} established conditions for when the number of
occurrences of a given subgraph in random graphs follows a normal
distribution.

\medskip

The plan of the paper is as follows.  In
Section~\ref{sec:countingtrees} we give a short introduction to
counting trees with generating functions, and also expand this to two
variables for counting stars (nodes of specific degree $k$) in
trees. In Section~\ref{sec:countingpatterns} we expand this framework
to the counting of patterns in trees. The resulting asymptotics are
presented in Section~\ref{sec:asymptoticbehaviour}, concluding the
proof of Theorem~\ref{Th1}.  Technical details
for this as well as explicit algorithms can be found in the appendix.
In fact, the algorithmic aspect is one of the driving forces of this
paper.

\section{Counting Trees and Counting Stars in Trees}
\label{sec:countingtrees}

In this section we introduce a three-step program to count the
number of trees in $\Tn{}$ and in the same fashion the number of
occurrences of nodes of degree $k$ in $\Tn{}$.  While redundant and
probably heavy in this simplistic situation, this procedure was
crucial to the derivation in~\cite{DrGi99} for counting stars and will
generalise well to our setting of general tree patterns.

For this purpose we make use of the sets \Rn{} of rooted labeled
trees of size $n$ and \Pn{} of planted labeled trees of size $n$.
For rooted and unrooted trees, the size~$n$ counts the total number
of nodes, whether internal or at the leaves.  On the other hand, a
planted tree is just a rooted tree where the root is adjoined an
additional ``phantom'' node which does not contribute to the size of
the tree, whereas the degree of the root is increased by one.  As
well, one can think of a planted tree as a rooted tree with an
additional edge having no end vertex. The advantage of using planted
trees, though it seems to add complexity, will be explained below.
Obviously $|\Pn{}|=|\Rn{}|$ and $|\Tn{}|=|\Rn{}|/n$. It is also well
known that $|\Rn{}| = n^{n-1}$ and $|\Tn{}|= n^{n-2}$.

The three-step program is the following one: First, the generating
function enumerating planted trees is determined, then it is used to
count rooted trees by deriving their generating function, and finally
the generating function counting unrooted trees is computed.

We define
\[
p(x) = \sum_{n=0}^\infty |\Pn| \frac{x^n}{n!},\quad
r(x) = \sum_{n=0}^\infty |\Rn| \frac{x^n}{n!},\quad
t(x) = \sum_{n=0}^\infty |\Tn| \frac{x^n}{n!}
\]
and proceed in the following way:
\begin{enumerate}
\item {\bf Planted Rooted Trees:}
A planted tree is a planted root node with zero, one, two, $\dots$
planted subtrees of any order.  In terms of the generating function
this yields
\begin{equation*}
p(x) = \sum_{n=0}^\infty \frac{x p(x)^n}{n!} = x e^{p(x)}.
\end{equation*}
\item {\bf Rooted Trees:}
For rooted trees we get the same (except for the phantom nodes which
are not present here), just a root with zero, one, two,
$\dots$ planted subtrees of any order
\begin{equation*}
r(x) = \sum_{n=0}^\infty \frac{x p(x)^n}{n!} = x e^{p(x)}  = p(x).
\end{equation*}
\item{\bf Unrooted Trees:}
Finally, we have $|\Tn{}|=|\Rn{}|/n$, as already mentioned. However,
we can also express $t(x)$ by a relation which
follows from a natural bijection between
rooted trees on the one hand and unrooted trees and pairs of planted rooted
trees (that are joined by identifying the additional edges at their planted
roots and discarding the phantom nodes) on the
other hand.\footnote{Consider the class of rooted (labeled) trees. If
the root is labeled by~$1$ then consider the tree as an unrooted
tree. If the root is not labeled by~$1$ then consider the first edge
of the path between the root and~$1$ and cut the tree into two planted
rooted trees at this edge.}  This yields
\begin{equation*}
t(x) = r(x) - \frac 12p(x)^2.
\end{equation*}
\end{enumerate}
The functional equation for $p(x)$ can be either used to extract
the explicit number $|\Pn|=n^{n-1}$ via Lagrange inversion or to obtain
the radius of convergence and asymptotic expansions of the singular
behaviour of this function. It is well known that $x_0 = 1/e$ is the common
radius
of convergence
of $p(x)$, $r(x)$, and $t(x)$, and that the singularity
at $x= x_0$ is of square-root type:
\begin{align*}
p(x) &= r(x) = 1- \sqrt 2 \sqrt{1-ex} + \frac 23(1-ex) + \cdots,\\
t(x) &= \frac 12 - (1-ex) + \frac{2\sqrt{2}}3 (1-ex)^{3/2} + \cdots.
\end{align*}
This is reflected by the asymptotic expansions of the numbers
\begin{align*}
|\Pn{}| &= |\Rn{}| = n^{n-1} \sim  \frac{n!}{\sqrt{2\pi}} e^n n^{-3/2},\\
|\Tn{}| &= n^{n-2} \sim \frac{n!}{\sqrt{2\pi}} e^n n^{-5/2}.
\end{align*}

In order to demonstrate the usefulness of the three-step procedure above
we repeat the same steps for
counting  stars with $k$ edges in trees, that is, the number of nodes of
degree $k$, a given fixed positive number.
Let $p_{n,m}$ denote the number of planted trees of size $n$
with exactly $m$ nodes of degree $k$. Furthermore, let $r_{n,m}$ and
$t_{n,m}$ be the corresponding numbers for rooted and unrooted trees and
set
\[
p(x,u) = \sum_{n,m=0}^\infty p_{n,m} \frac{x^nu^m}{n!},\quad
r(x,u) = \sum_{n,m=0}^\infty r_{n,m} \frac{ x^nu^m}{n!},\quad
t(x,u) = \sum_{n,m=0}^\infty t_{n,m} \frac{ x^nu^m}{n!}.
\]
Then we have (compare with \cite{DrGi99})
\begin{enumerate}
\item {\bf Planted Rooted Trees:}
\begin{equation*}\label{eq:startrees}
 p(x,u)=\sum_{\substack{n=0\\n\neq
k-1}}^\infty \frac{x p(x,u)^n}{n!} +
\frac{xup(x,u)^{k-1}}{(k-1)!} =x e^{p(x,u)} +
\frac{x(u-1)p(x,u)^{k-1}}{(k-1)!}.
\end{equation*}
\item {\bf Rooted Trees:}
\begin{equation*}
r(x,u)=\sum_{\substack{n=0\\n\neq k}}^\infty \frac{x
p(x,u)^n}{n!} + \frac{xup(x,u)^k}{k!}
= x e^{p(x,u)}+
\frac{x(u-1)p(x,u)^k}{k!}.
\end{equation*}
\item {\bf Unrooted Trees:}
Similarly to the above we have $t_{n,m}=r_{n,m}/n$ which is
sufficient for our purposes. However, as above,
it is  also possible to express $t(x,u)$ by
\begin{equation*}
t(x,u)= r(x,u) - \frac 12 p(x,u)^2.
\end{equation*}
\end{enumerate}
Note that the use of the notion of planted trees is crucial in order
to keep track of the nodes of degree $k$ by means of the recursive
structure of planted trees.
In \cite{DrGi99} this approach was used to show that the asymptotic
distribution of the number of nodes of degree $k$ in trees of size $n$ is
normal,
with expectation and variance proportional to $n$.

\section{Counting Patterns in Trees}
\label{sec:countingpatterns}

\begin{figure}[htb]
\centering
\includegraphics[width=6cm]{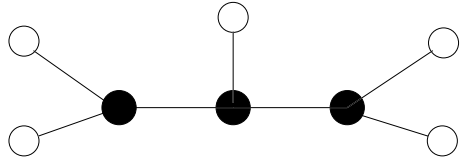}
\caption{Example pattern} \label{fig:pattern}
\end{figure}

We now generalize the counting procedure of
Section~\ref{sec:countingtrees} to more complicated
patterns. For our purpose, a pattern is a given (finite unrooted unlabeled)
tree \Mk{}. To ease explanations, we will use as \Mk{} the example
graph in Figure~\ref{fig:pattern}.

We say that a specific pattern
\Mk{} occurs in a tree~$T$ if \Mk{} occurs in~$T$ as an induced
subtree in the sense that the node degrees for the internal
(filled) nodes in the pattern match the degrees of the corresponding nodes
in~$T$, while the external (empty) nodes match nodes of
arbitrary degree.\footnote {More generally we could also consider
pattern-matching problems
for patterns in which some degrees of certain possibly external ``filled'' nodes
must
match exactly while the degrees of the other, possibly internal ``empty'' nodes
might be
different. But then the situation is more involved, see
Section~\ref{sec:extensions}.} Because the results for the patterns
consisting of only one node or two nodes and one edge are trivial,
we now concentrate on patterns with at least three nodes.

Our principal aim is to get relations for the generating functions which  count
the number of occurrences of a specific pattern \Mk{}.
Let $p_{n,m}$ denote the number of planted rooted trees with
$n$~nodes and exactly $m$~occurrences of the pattern $\Mk{}$ and let
\[
p = p(x,u) = \sum_{n,m=0}^\infty  p_{n,m}\frac{x^n u^m}{n!}
\]
be the corresponding generating function.

\subsection{Generating Functions for Planted Rooted Trees}

\begin{proposition}\label{Pro1} ({\bf Planted Rooted Trees})
Let \Mk{} be a pattern. Then there exists a certain number $L+1$ of
auxiliary functions $a_j(x,u)$ $(0\le j\le L)$ with
\[
p(x,u) = \sum_{j=0}^{L} a_j(x,u)
\]
and polynomials $P_j(y_0,\ldots,y_{L},u)$  $(1\le j\le L)$ with
non-negative coefficients such that
\begin{equation}
\begin{aligned}
a_0(x,u) &= xe^{a_0(x,u)+ \cdots + a_L(x,u)} - x \sum_{j=1}^{L}
P_j(a_0(x,u),\ldots,a_L(x,u),1) \\
a_1(x,u) &= x\cdot P_1(a_0(x,u),\ldots,a_L(x,u),u)\\
&\ \vdots \\
a_L(x,u) &=  x\cdot P_{L}(a_0(x,u),\ldots,a_L(x,u), u).
\label{eq:asystem}
\end{aligned}
\end{equation}
Furthermore,
\[
\sum_{j=1}^{L}  P_j(y_0,\ldots,y_{L},1) \le_c e^{y_0+\cdots+y_{L}},
\]
where $f \le_c g$ means that all Taylor coefficients of the left-hand
side are smaller than or equal to
the corresponding coefficients of the right-hand side.
Moreover, the dependency graph of this system is strongly
connected.\footnote {The notion of dependency graph is explained
in Appendix~\ref{sec-asymptotics} and intuitively speaking, reflects
the fact that no
subsystem can be solved before the whole system.}
\end{proposition}

The proof of this proposition is in fact the core of the paper. In
order to make the arguments more transparent we will demonstrate
them with the help of the example pattern in Figure~\ref{fig:pattern}. At
each step of the proof we will also indicate how to make all
constructions explicit so that it is possible to generate
System~(\ref{eq:asystem}) effectively.

\begin{figure}
\centering
\includegraphics[width=8cm]{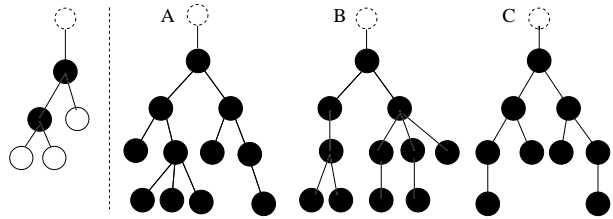}
\caption{Planted pattern matching}
\label{fig:plantedmatch}
\end{figure}

In a first step we introduce the notion of a
{\em planted pattern}. A planted pattern \Mp\ is just a planted
rooted tree where we again distinguish between internal (filled) and
external (empty) nodes. It matches a planted rooted tree from~\Tn\ if
\Mp\ occurs as an induced subtree starting from the (planted) root,
that is, the branch structure and node degrees of the filled nodes
match. Two occurrences may overlap. For example, in
Figure~\ref{fig:plantedmatch} the planted pattern~\Mp\ on the left
matches the planted tree $A$ twice (following the left, resp.\ the
right edge from the root), but $B$ not at all.  Also remark that,
notwithstanding the symmetry of~$C$, the pattern~\Mp\ really matches $C$
twice, as we are interested in matches in labeled trees.

\begin{figure}
\centering
\includegraphics[width=6cm]{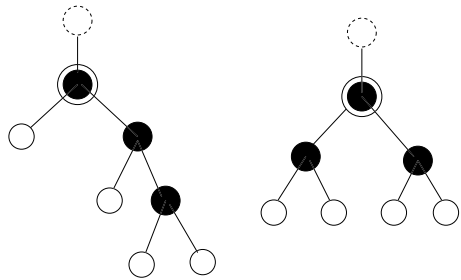}
\caption{Planted patterns for the pattern in Figure~\ref{fig:pattern}}
\label{fig:plantedpatterns}
\end{figure}

We now construct a {planted pattern} for each internal (filled) node
of our pattern \Mk{} which is adjacent to an external (empty) node.
The internal (filled) node is considered as the planted root and one
of the free attached leaves as the plant. In our example we obtain
the two graphs in Figure~\ref{fig:plantedpatterns}.

\begin{figure}[b]
\centering
\includegraphics[width=10cm]{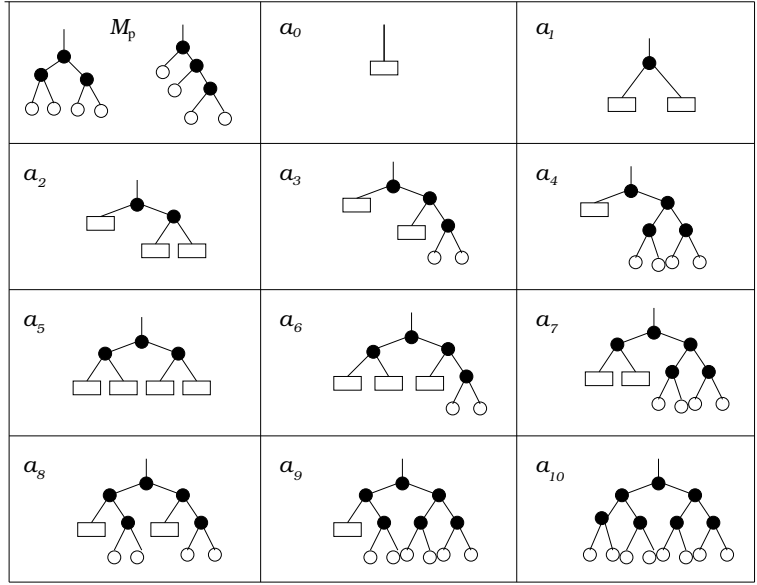}
\caption{Tree partition} \label{fig:tree-partition}
\end{figure}

The next step is to partition all planted trees according to their
degree distribution up to some adequate level.  To this end, let $D$
denote the set of out-degrees that occur in the planted patterns
introduced above and $h$~be the maximal height of these patterns.  In our
example we have $D =\{2\}$ and $h = 3$.  For obtaining a partition, we
more precisely consider all trees of height less than or equal to~$h$
with out-degrees in~$D$. We distinguish two types of leaves in these trees,
depending on the depth at which they appear:
leaves in level~$h$, denoted ``$\circ$'', and leaves at levels
less than~$h$, denoted ``$\Box$''.  For our example we get 11~different
trees $a_0, a_1, \ldots, a_{10}$, depicted on Figure~\ref{fig:tree-partition}.

These trees induce a natural partition of all planted trees for the
following interpretation of the two types of leaves:  We say that a
tree $T$ is contained in class\footnote{By abuse of notation the tree
class corresponding to the finite tree $a_j$ is denoted by the same
symbol~$a_j$.} $a_j$ if it matches the finite tree (or pattern) $a_j$
in such a way that a node of type~$\Box$ has degree not in~$D$, while
a node of type~$\circ$ has any degree. For example,
$a_0$~corresponds to those planted trees where the out-degree of the
root is not in $D$.

It is easy to observe that these (obviously disjoint) classes of
trees form a partition. Indeed, take any rooted tree.  For any path
from the root to a leaf, consider the first node with out-degree not
in~$D$, and replace the whole subtree at it with~$\Box$.  Then
replace any node at depth~$h$ with~$\circ$.  The tree obtained in
this way is one in the list.

Furthermore, the classes above can be described
recursively.  To this end, it proves
convenient to introduce a formal notation to describe operations
between classes of trees: $\oplus$~denotes the disjoint union of
classes; $\setminus$~denotes set difference; recursive descriptions of tree
classes are given in the form $a_i=\nc a_{j_1}^{e_1}\dotsm
a_{j_\ell}^{e_\ell}$, to express that the class~$a_i$ is constructed
by attaching $e_1$~subtrees from the class~$a_{j_1}$, $e_2$~subtrees
from the class~$a_{j_2}$, etc, to a root node that we denote~$\nc$.

In our example we get the following relations:
\begin{align*}
a_0 &= p \setminus \bigoplus_{i=1}^{10} a_i = x \oplus
x\bigoplus_{i=0}^{10} a_i \oplus  x\bigoplus_{n=3}^\infty \biggl(
\bigoplus_{i=0}^{10} a_i\biggr)^n,\\
a_1 &= \nc a_0^2, \\
a_2 &= \nc a_0 a_1,\\
a_3 &= \nc a_0(a_2 \oplus a_3\oplus a_4),\\
a_4 &= \nc a_0(a_{5} \oplus a_6\oplus a_7\oplus a_8\oplus a_9\oplus a_{10}),\\
a_5 &= \nc a_1^2, \\
a_6 &= \nc a_1(a_2 \oplus a_3\oplus a_4),\\
a_7 &= \nc a_1(a_{5} \oplus a_6\oplus a_7\oplus a_8\oplus a_9\oplus a_{10}),\\
a_8 &= \nc(a_2 \oplus a_3\oplus a_4)^2,\\
a_{9}&= \nc(a_2 \oplus a_3\oplus a_4) (a_{5} \oplus a_6\oplus a_7\oplus a_8\oplus a_9\oplus a_{10}),\\
a_{10} &= \nc(a_{5} \oplus a_6\oplus a_7\oplus a_8\oplus a_9\oplus a_{10})^2.
\end{align*}
This is to be interpreted as follows.  Trees in~$a_1$ consist of a
(planted) root that is denoted by~$\nc$ that has out-degree~$2$, and two
children that are of out-degree distinct from~$2$, that is, in~$a_0$.
Similarly, trees in~$a_3$ consist of a root~$\nc$ with out-degree $2$
and subject to the following additional constraints: one subtree at the root is
exactly of
type~$a_0$; the other subtree, call it~$T$, is of out-degree~2,
either with both subtrees
of degree other than~$2$ (leading to $T$~in~$a_2$), or with one subtree of degree~$2$ and the other of
degree other than~$2$ (leading to $T$~in~$a_3$), or 
with both of its subtrees of degree~$2$ (leading to $T$~in
class~$a_4$). Summarizing:
$a_3 = \nc a_0(a_2 \oplus a_3\oplus a_4)$.  Of course
this can be also interpreted as $a_3 = \nc a_0a_2 \oplus \nc a_0a_3
\oplus \nc a_0a_4$.
Another more involved example corresponds to $a_8$; here both subtrees
are of the form $a_2 \oplus a_3\oplus a_4$.

To show that the recursive description can be obtained easily in
general, consider a tree~$a_j$ obtained from some planted
pattern~$\Mp$.  Let $s_1$, \dots, $s_d$ denote its subtrees at the
root.  Then, in each~$s_i$, leaves of type~$\circ$ can appear only
at level~$h-1$. Substitute for all such~$\circ$ either~$\Box$ or a
node of out-degree chosen from~$D$ and having~$\circ$ for all its
subtrees. Do this substitution in all possible ways.  The collection
of trees obtained are some of the~$a_k$'s, say $a_{k^{(j)}_1}$,
$a_{k^{(j)}_2}$, etc.  Thus, we obtain the recursive relation
$a_j=x(a_{k^{(1)}_1}\oplus
a_{k^{(1)}_2}\oplus\dotsb)\dotsm(a_{k^{(d)}_1}\oplus
a_{k^{(d)}_2}\oplus\dotsb)$ for~$a_j$.

In general, we obtain a partition of $L+1$ classes
$a_0,\ldots,a_L$ and corresponding recursive descriptions, where
each tree type $a_j$ can be expressed as a disjoint union of tree
classes of the kind
\begin{equation}\label{eq:configuration}
\nc a_{j_1} \dotsm a_{j_r}=\nc a_0^{l_0}\dotsm a_L^{l_L},
\end{equation}
where $r$ denotes the degree of the root of $a_j$ and the non-negative
integer~$l_i$ is the number of repetitions of the tree type~$a_i$.

We proceed to show that this directly leads to a system of
equations of the form (\ref{eq:asystem}), where
each polynomial relation stems from a recursive equation between
combinatorial classes.

Let $\Lambda_j$ be the set of tuples $(l_0,\ldots,l_{L})$ with the property that
$(l_0,\ldots,l_{L}) \in \Lambda_j$ if and only if the term of type
(\ref{eq:configuration}) is involved in the recursive description
of~$a_j$ (in expanded form).  Further, let $k=K(l_0,\ldots,l_{L})$
denote the number of {\em additional occurrences\/} of the pattern
$\Mk{}$ in (\ref{eq:configuration}) in the following sense: if
$b=\nc a_{j_1} \dotsm a_{j_{r}}$ and $T$ is a (planted rooted) labeled tree of $b$
with subtrees $T_1\in a_{j_1}$, $T_2\in a_{j_2}$, etc, and
$\Mk{}$~occurs $m_1$~times in~$T_1$, $m_2$~times in~$T_2$, etc, then
$T$ contains $\Mk{}$ exactly $m_1+ m_2+\cdots+ m_d + k$ times. The
number~$k$ corresponds to the number of occurrences of $\Mk{}$ in
$T$ in which the root of $T$ occurs as internal node of the pattern.
By construction of the classes $a_i$ this number only depends on $b$ 
and not on the particular tree $T\in
b$. Let us clarify the calculation of $k=K(l_0,\dots,l_L)$ with an
example. Consider the class $a_{9}$ of the partition for the example pattern. Now, in order to determine the number of additional occurrences, we match the planted
patterns of Figure~\ref{fig:plantedpatterns} at the root of an
arbitrary tree of class $a_{9}$. The left planted pattern of
Figure~\ref{fig:plantedpatterns} matches three times, the right one
matches once. Thus we find that in this case $k=4$.
For the other classes we find the following values of
$k=K(l_0,\dots,l_L)$:
\begin{center}
 \begin{tabular}{l||c|c|c|c|c|c|c|c|c|c|c}
 Terms of class & $a_0$ & $a_1$ & $a_2$ & $a_3$ & $a_4$ & $a_5$ & $a_6$ &
 $a_7$ & $a_8$ & $a_9$ & $a_{10}$ \\ \hline
 Value of $k$ & 0 & 0 & 0 & 1 & 2 & 1 & 2 & 3 & 3 & 4 & 5
 \end{tabular}.
\end{center}
\medskip

Now define series~$P_j$ by
\[
P_j(y_0,\ldots,y_{L},u) = \sum_{(l_0,\ldots,l_{L})\in \Lambda_j}
\frac{1} {l_0!\cdots l_{L}!} y_0^{l_0} \cdots y_{L}^{l_{L}}
u^{K(l_0,\ldots,l_{L})}.
\]
These are in fact polynomials for $1\le j\le L$ by the finiteness of
the corresponding~$\Lambda_j$. All matches of the planted
patterns are handled in the $P_j$, $1\le j\le L$, thus
\[
P_0(y_0,\ldots,y_{L},u) = e^{y_0+\cdots+y_{L}} - \sum_{j=1}^{L}
P_j(y_0,\ldots,y_{L},1)
\]
does not depend on $u$.

In our pattern we get for example for $P_8(y_0,\dots, y_{10},u)$
\[
P_8(y_0,\dots, y_{10},u) =  \frac 12 xy_2^2u^3 +xy_2y_3u^3 +  xy_2y_4u^3+\frac 12 xy_3^2u^3 +   
xy_3y_4u^3 + \frac 12 xy_4^2u^3  = \frac 12 x(y_2+y_3+y_4)^2u^3.
\]
Finally, let $a_{j;n,m}$ denote the number of planted rooted trees
of type $a_j$ with $n$ nodes and $m$~occurrences of the pattern $\Mk{}$
and set
\begin{equation*}
a_j(x,u)=\sum_{n,m=0}^\infty  a_{j;n,m}\frac{x^n u^m}{n!}.
\end{equation*}
By this definition it is clear that
\[
a_j(x,u) = x\cdot P_j\bigl(a_0(x,u),\ldots,a_L(x,u),u\bigr),
\]
because the size of labeled trees is counted by~$x$ (exponential generating function) 
and the occurrences of the patterns is additive and counted by~$u$.
Hence, we explicitly obtain the proposed structure of the system of functional
equations (\ref{eq:asystem}).

For the example pattern we arrive at the following system of
equations, where we denote the generating function of the
class~$a_i$ by the same symbol~$a_i$:
\begin{align*}
a_0 &=  a_0(x,u) = p - \sum_{i=1}^{10} a_i = x + x
\sum_{i=0}^{10} a_i +  x\sum_{n=3}^\infty \frac 1{n!} \left(
\sum_{i=0}^{10} a_i\right)^n,\\
a_1 &= a_1(x,u) = \frac 12 x a_0^2, \\
a_2 &= a_2(x,u) = x a_0  a_1 ,\\
a_3 &= a_3(x,u) =  x a_0  (a_2 + a_3+ a_4) u ,\\
a_4 &= a_4(x,u) = x a_0   (a_{5} + a_6+ a_7+ a_8+
a_9+ a_{10}) u^2,\\
a_5 &= a_5(x,u) =  \frac 12 x a_1^2 u,\\
a_6 &= a_6(x,u) = x a_1  (a_2 + a_3+ a_4) u^2 ,\\
a_7 &= a_7(x,u) = x a_1   (a_{5} + a_6+ a_7+ a_8+
a_9+ a_{10}) u^3,\\
a_8 &= a_8(x,u) =  \frac 12 x(a_2 + a_3+ a_4)^2 u^3,\\
a_{9} &= a_{9}(x,u) =x(a_2 + a_3+ a_4)  (a_{5} + a_6+ a_7+ a_8+
a_9+ a_{10}) u^4,\\
a_{10} &= a_{10}(x,u) = \frac 12  x(a_{5} + a_6+ a_7+ a_8+
a_9+ a_{10})^2 u^5.
\end{align*}

In order to complete the proof of Proposition~\ref{Pro1} we just
have to show that the dependency graph is strongly connected. By
construction, $a_0=a_0(x,u)$ depends on all functions
$a_i=a_i(x,u)$. Thus, it is sufficient to prove that every $a_i$
($1\le i \le L$) also depends on~$a_0$. For this purpose consider
the subtree of \Mk{} that was labeled by $a_i$ and consider a path
from its root to an empty node. Each edge of this path corresponds
to another subtree of \Mk{}, say $a_{i_2}$, $a_{i_3}, \ldots,
a_{i_r}$. Then, by construction of the system of functional
equations above, $a_i$ depends on $a_{i_2}$, $a_{i_2}$ depends on
$a_{i_3}$ etc. Finally the root of $a_{i_r}$ is adjacent to an empty
node and thus (the corresponding generating function) depends on
$a_0$. This completes the proof of Proposition~\ref{Pro1}.

\bigskip

\begin{figure}[htb]
\centering
\includegraphics[width=10cm]{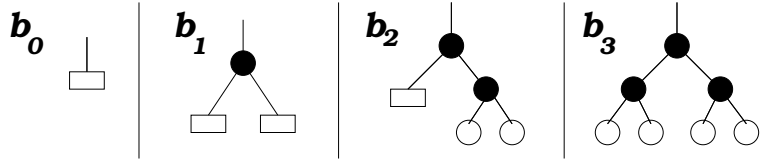}
\caption{The classes corresponding to the $b_i$ of equations \eqref{eq:bclasses}}
\label{fig:bclasses}
\end{figure}

Note that we obtain a relatively more compact form of this system by introducing
\begin{equation}
\begin{aligned}\label{eq:bclasses}
b_0 &= b_0(x,u) = a_0(x,u), \\
b_1 &= b_1(x,u) = a_1(x,u), \\
b_2 &= b_2(x,u) = a_2(x,u) + a_3(x,u) + a_4(x,u) \\
b_3 &= b_3(x,u) = a_{5}(x,u) + a_6(x,u) + a_7(x,u) + a_8(x,u) + a_9(x,u)
+a_{10}(x,u),
\end{aligned}
\end{equation}
together with the recursive relations
\begin{equation*}
\begin{aligned}
b_0 &=xe^{b_0+b_1+b_2+b_3} -\frac12 x(b_0+b_1+b_2+b_3)^2, \\
b_1 &=\frac12xb_0^2, \\
b_2 &=xb_0b_1+xb_0b_2u+xb_0b_3u^2, \\
b_3 &= \frac12 xb_1^2u + xb_1b_2u^2 +  xb_1b_3u^3 + \frac12 xb_2^2u^3  +
 +xb_2b_3u^4+ \frac12xb_3^2u^5 .
\end{aligned}
\end{equation*}
The combinatorial classes corresponding to the $b_i$ (which we will
also denote by $b_i$) have the interpretation shown in Figure
\ref{fig:bclasses}. We could have obtained the classes~$b_i$ directly by restraining the
construction to a maximal depth~$h-1$ instead of~$h$. In principle,
we could then apply the analytic treatment of
Section~\ref{sec:asymptoticbehaviour} to the system of the~$b_i$.
However we feel that the existence of a recursive structure of the
system of the~$b_i$ with a well-defined $K(l_0,.., l_L)$ for each
term in the recursive description is slightly less clear. Therefore
we preferred to work with the~$a_i$ which have a
well-defined~$K(a_i)$. In Appendix~\ref{sec-algorithms} we will
discuss another algorithm that yields in general even more compact
systems of equations.

\subsection{From Planted Rooted Trees to Rooted and Unrooted Trees}

The next step is to find equations for the exponential generating
function of rooted trees (where occurrences of the pattern are marked
with~$u$). As above we set
\[
r(x,u) = \sum_{n,m=0}^\infty r_{n,m} \frac{x^n u^m}{n!},
\]
where $r_{n,m}$ denotes the number of rooted trees of size~$n$ with
exactly $m$~occurrences of the pattern~\Mk. (That is, occurrences
of the \emph{rooted} patterns~$\Mr$ deducible from~$\Mk$. Here, a
rooted pattern is defined in a very similar way as a planted
pattern.)

\begin{proposition}\label{Pro2} ({\bf Rooted Trees})
Let \Mk{} be a pattern
and let
\[
a_0(x,u),\ldots,a_L(x,u)
\]
denote the auxiliary functions introduced in Proposition~\ref{Pro1}.
Then there exists a polynomial \linebreak $Q(y_0,\ldots,y_{L},u)$
with non-negative coefficients satisfying $Q(y_0,\ldots,y_{L},1)\le_c
e^{y_0+\cdots+y_{L}}$, and such that
\begin{equation}\label{eqPro21}
r(x,u) = G(x,u,a_0(x,u),\ldots,a_L(x,u))
\end{equation}
for
\begin{equation}\label{eqPro22}
G(x,u,y_0,\ldots,y_{L})=
x\left(e^{y_0+\dots+y_L} - Q(y_0,\ldots,y_{L},1) + Q(y_0,\ldots,y_{L},u)\right).
\end{equation}
\end{proposition}

\begin{proof}
The proof is in principle a direct continuation of the proof of
Proposition \ref{Pro1}. We recall that a rooted tree is just a root
with zero, one, two, $\dots$ planted subtrees, i.e., the class of
rooted trees can be described as a disjoint union of classes~$c$ of
rooted trees of the form $\nc a_{j_1} \dotsm a_{j_d}$. Furthermore, let
$l_i$ denote the number of classes $a_i$ in this term such that $c=\nc
a_0^{l_0}\dotsm a_L^{l_L}$, and set $\bar K(l_0,\ldots,l_L)$ to be
the number of additional occurrences of the pattern $\Mk{}$. This
number again corresponds to the number of occurrences of~$\Mk$ in a
(rooted) tree $T\in c$ in which the root of $T$ occurs as internal node of
the pattern.
Set
\[
Q_d(y_0,\ldots,y_{L},u) = \sum_{l_0+\dots+l_L =d} \frac{1}
{l_0!\cdots l_{L}!} y_0^{l_0} \cdots y_L^{l_{L}}
u^{\bar K(l_0,\ldots,l_{L})}.
\]
Then by construction
\[
r(x,u) = x \sum_{d\ge 0} Q_d(a_0(x,u),\ldots,a_L(x,u),u).
\]
Note that $\sum_{d\ge 0} Q_d(y_0,\ldots,y_{L},1) =
e^{y_0+\cdots+y_{L}}$. Let $\bar D$ denote the set of degrees of the
internal (filled) nodes of the pattern, that is, $\bar D=\{\,d+1:d\in D\,\}$; then
$Q_d(y_0,\ldots,y_{L},u)$ does not depend on $u$ if $d\not\in\bar D$. With
\[
Q(y_0,\ldots,y_{L},u) := \sum_{d\in\bar D} Q_d(y_0,\ldots,y_{L},u),
\]
we obtain (\ref{eqPro21}) and (\ref{eqPro22}). The number~$\bar
K(l_0,\dots,l_L)$ is well-defined for a similar reason as was
$K(l_0,\dots,l_L)$, and can be calculated similarly.
\end{proof}

\begin{figure}[htb]
\centering
\includegraphics[width=6cm]{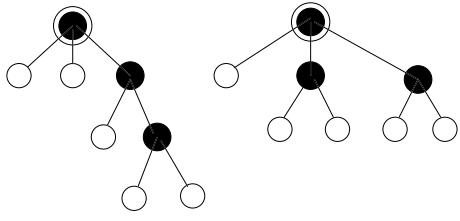}
\caption{Rooted patterns for the pattern in Figure~\ref{fig:pattern}}
\label{fig:rootpattern}
\end{figure}

We again illustrate the proof with our example. In
Figure~\ref{fig:rootpattern} the corresponding rooted patterns are
shown. For convenience let $r_0=r_0(x,u)$ denote the function
\[
r_0 = xe^p-  \frac{xp^3}{3!},
\]
where $p = a_0+\cdots+a_{10}$.  The function~$r_0$ might also be
interpreted as a
catch-all function for the ``uninteresting'' subtrees---just a
root~$\nc$ with an unspecified number of planted trees attached,
except the ones we handle differently, namely the cases $d\in\bar D = \{
3 \}$. The generating function $r=r(x,u)$ for rooted trees is then
given by
\begin{equation*}
r = r_0
 + \frac 16 x b_0^3 + \frac 12 x\sum_{1\le i \le 3} b_0^2 b_i u^{i-1} + \frac 12 x\sum_{1\le
 i,j \le 3} b_0 b_i b_j u^{i+j-1} + \frac 16 x \sum_{1\le i,j,k \le 3}
  b_i b_j b_k u^{i+j+k}
\end{equation*}
where the~$b_i$ are defined in \eqref{eq:bclasses}.

As above we have $t_{n,m}=r_{n,m}/n$, where $t_{n,m}$ denotes the
number of unrooted trees with $n$~nodes and exactly $m$ occurrences
of the pattern \Mk. This relation is sufficient for our purposes. It
is also possible to express the corresponding generating function
$t(x,u)$. In a way similar as before, we can define the number of
additional occurrences $\hat K(i,j)$ of the pattern \Mk{} that
appear by constructing an unrooted tree from two planted trees of
the class $a_i$ and $a_j$ by identifying the additional edges at
their planted roots and discarding the phantom nodes. For our
example we get 
\begin{equation*}
t(x,u) = r(x,u) - \frac 12 p(x,u)^2  - \frac 12 \sum_{1\le
i,j \le 3} b_i(x,u) b_j(x,u) (u^{i+j-2} -1).
\end{equation*}

\section{Asymptotic Behavior}
\label{sec:asymptoticbehaviour}

Since we are not interested in the actual number of occurrences of
the pattern, but only in its asymptotic behavior, we do not have to
compute explicit formulae from the system of equations.  Instead, we
apply a result slightly adapted from \cite{Dr97} which we state and
discuss in Appendix~\ref{sec-asymptotics}. In fact, it is
immediately clear that Theorem~\ref{th:syseqns} in this appendix,
whose object is the proof of Gaussian limiting distributions,
applies to the kind of problem we are interested in: the assertions
of Propositions~\ref{Pro1} and \ref{Pro2} exactly fit the
assumptions of Theorem~\ref{th:syseqns}.

The only missing point is the existence of a non-negative solution
$(x_0,\aaa_0)$ of the system
\begin{align}
\aaa &= \FF(x,\aaa,{ 1}),  \label{eq:F1}\\
0 &= \det(\II - \FF_\aaa(x,\aaa,{ 1})), \label{eq:F2}
\end{align}
where \eqref{eq:F1}~is the system of functional equations
of Proposition~\ref{Pro1}
and $\FF_\aaa$~is the Jacobian matrix of~$\FF$.
Since the sum of all unknown functions $p(x,u)$ is known for
$u=1$:
\[
p(x,1) = p(x) = \sum_{n\ge 1} n^{n-1} \frac{x^n}{n!} = 1 - \sqrt 2
\sqrt{1-ex} + \cdots,
\]
it is not unexpected that $x_0=1/e$.

\begin{proposition}\label{Pro3}
There exists a unique non-negative solution $(x_0,\aaa_0)$ of
System~(\ref{eq:F1}--\ref{eq:F2}), for which $x_0=1/e$ and the components of $\aaa_0$ 
are polynomials (with rational coefficients) in $1/e$.
\end{proposition}

\begin{proof}
For a proof, set $u=1$ and consider the solution $\aaa(x,1) =
(a_0(x,1),\ldots, a_{L-1}(x,1))$. Since the dependency graph is
strongly connected it follows that all functions $a_j(x,1)$ have the
same radius of convergence which has to be $x_0=1/e$, and all
functions are singular at $x=x_0$. Since $0\le a_j(x,1)\le p(x,1) <
\infty$ for $0\le x \le x_0$ it also follows that $a_j(x_0,1)$ is
finite, and we have $\aaa(x_0,1) = \FF(x_0,\aaa(x_0,1),1)$. If we
had the inequality $\det (\II - \FF_\aaa(x_0,\aaa(x_0,1),{ 1}))\ne
0$ then the implicit function theorem would imply the existence of
an analytic continuation for $a_j(x,1)$ around $x=x_0$, which is, of
course, a contradiction. Thus, the determinant is zero and system~(\ref{eq:F1}--\ref{eq:F2}) has a unique solution.

To see that the components $\bar a_0,\dots, \bar a_L$ (with $\bar a_i = a_i(1/e,1)$) of 
$\aaa_0$ are polynomials in $1/e$ we will construct the partition $\mathcal A=\{a_0,a_1,\dots,a_L\}$ 
on which the system of equations (\ref{eq:F1}--\ref{eq:F2}) is based by refining step by step the 
trivial partition consisting of only one class $p$.  The recursive description of this trivial 
partition is given by the formal equation $p=x \sum_{i\ge 0} p^i$.
Additionally, the solution of the corresponding equation $p=x\exp(p)$ for the generating function $p$ 
(denoted by the same symbol $p$) is given by $(x_0,\bar p)=(1/e,1)$, with $\bar p$ clearly a 
(constant) polynomial in $1/e$. Now let $D=\{d_1,\dots,d_s\}\ (s\in\mathbb N)$ again denote 
the set of out-degrees that occur in the planted patterns. We will refine $p$ by introducing 
for each $d_i\in D$ a class $a_{i}$ consisting of all trees of root out-degree~$d_i$, as well as a class~$a_0$ for trees with root out-degree not in $D$. The partition 
$\{a_0, a_1,\dots, a_s\}$ has the recursive description 
\begin{align}
a_0 &= x\sum_{j\in \mathbb N \setminus D}(a_0 \oplus a_1 \oplus \cdots \oplus a_s)^j, \nonumber \\
a_i &=x (a_0\oplus a_1 \oplus \cdots \oplus a_s)^{d_i} \qquad (i=1,\dots, s), \label{eq:pol}
\end{align}
and the solution of the corresponding system of equations
\begin{align}
a_0(x,1) &= x\sum_{j\in \mathbb N \setminus D}\frac 1{j!} (a_0(x,1)+ a_1(x,1)+ \cdots + a_s(x,1))^j 
 \nonumber \\
&= xe^{a_0(x,1)+\cdots + a_s(x,1)}
 - x\sum_{i=1}^s \frac 1{d_i!} (a_0(x,1)+ a_1(x,1)+ \cdots + a_s(x,1))^{d_i} \nonumber \\
 &= xe^{p(x)}
 - x\sum_{i=1}^s \frac 1{d_i!} p(x)^{d_i},
 \label{eq:pol-2} \\
a_i(x,1)  &=\frac{x}{d_i!} (a_0(x,1)+ a_1(x,1)+ \cdots + a_s(x,1))^{d_i} 
= \frac{x}{d_i!} p(x)^{d_i} \qquad (i=1,\dots, s), \nonumber
\end{align}
is given by
\begin{equation}\label{barai}
x_0=1/e,\qquad \bar a_i = \frac1{d_i!\,e}\quad (i=1,\dots ,s),\qquad 
\bar a_0 = 1- (\bar a_1 +\cdots+\bar a_s),
\end{equation}
 thus again polynomials in $1/e$. We continue by refining this last partition by introducing 
classes $c_1,\dots,c_m$ (for some $m\in\mathbb N$) for each term at the right-hand side of 
\eqref{eq:pol} after expanding the ``multinomial''. Such a class $c_j$ is of the form 
$c_j=xa_0^{l_0^{(j)}} a_1^{l_1^{(j)}}\cdots a_s^{l_s^{(j)}}$ with 
natural numbers $l_i^{(j)},\ i=0,\dots, s$. We get a 
new partition $\{a_0, c_1,\dots,c_m \}$ which has a recursive description by 
construction (because we can replace the $a_i$ by disjoint unions of certain $c_j$). 
The corresponding system of equations for the generating functions is given by 
\[
c_j(x,1) =  \frac x{l_0^{(j)}!\,l_1^{(j)}!\cdots l_s^{(j)}!}
a_0(x,1)^{l_0^{(j)}}a_1(x,1)^{l_1^{(j)}}\cdots a_s(x,1)^{l_s^{(j)}}
\quad (j=1,\dots, u)
\]
and consequently we have for $x_0 = 1/e$ the solution
\[
\bar c_j = \frac 1e \frac1{l_0^{(j)}!\,l_1^{(j)}!\cdots l_s^{(j)}!}
\bar a_0^{l_0^{(j)}}\bar a_1^{l_1^{(j)}}\cdots \bar a_s^{l_s^{(j)}}
\quad (j=1,\dots, m)
\]
with the $\bar a_i$ of \eqref{barai}. Thus the $\bar c_j$ are again polynomials in $1/e$. 
By continuing this procedure until level $h$ (i.e., performing the refinement step $h$ times) 
we end up with the partition $\mathcal A$ and we see that the solution for the corresponding 
system of equations consists of polynomials in $1/e$, which completes
the proof of Proposition \ref{Pro3}.
\end{proof}

\medskip

Note that there is a close link with Galton--Watson branching processes. 
Let $p_k=\frac 1{k!\,e}$ denote a Poisson offspring distribution. Now we interpret a class $a_i$
as the class of process realizations for which the (non-planar) branching structure at the 
beginning of the processes corresponds to the root structure of $a_i$. 
Then $\bar a_i = a_i(1/e,1)$ is just the probability of this event.

\medskip

We now solve the system of equations obtained for the example pattern. 
We have $x_0 = 1/e$.  The components of~$\aaa_0$ can easily be obtained by following the
construction of the proof of Proposition~\ref{Pro4} (or we use
the branching process interpretation). For example, if we set $p = 1/(2e)$ for the
probability of an out-degree $2$ and $q = 1-p$ then we get 
$\bar a_4= a_4(1/e,1)= 2 qp^3= \frac{2e-1}{16e^5}$. The factor~$2$ 
comes from the fact that the two subtrees of the root may be interchanged, 
see Figure~\ref{fig:tree-partition}. The other classes can be treated similarly and we find:
\begin{equation}
\begin{aligned}\label{eq:sol}
 p(1/e, 1)   &=  1 ,                & a_5(1/e, 1)  &=  {(2e-1)^4}/{(128e^7)} ,\\
 a_0(1/e, 1) & =  {(2e-1)}/{(2e)},  & a_6(1/e, 1) &=  {(2e-1)^3}/{(32e^7)} ,\\
 a_1(1/e, 1)&  =   {(2e-1)^2}/{(8e^3)},  & a_7(1/e, 1) &=  {(2e-1)^2}/{(64e^7)} ,\\
a_2(1/e, 1) & =   {(2e-1)^3}/{(16e^5)}, & a_8(1/e, 1)  &=   {(2e-1)^2}/{(32e^7)} ,\\
  a_3(1/e, 1)  &=   {(2e-1)^2}/{(8e^5)}, & a_{9}(1/e, 1) &= {(2e-1)}/{(32e^7)},\\
 a_4(1/e, 1) & =  {(2e-1)}/{(16e^5)}, & a_{10}(1/e, 1)  &=   {1}/{(128e^7)} .
\end{aligned}
\end{equation}

\medskip

We are now ready to complete the proof of the main part of Theorem~\ref{Th1}.
By Propositions~\ref{Pro1}--\ref{Pro3} we can apply 
Theorem~\ref{th:syseqns} and it follows that the numbers
$r_{n,m}$ have a Gaussian limiting
distribution with mean and variance which are proportional to $n$.
Since $t_{n,m} = r_{n,m}/n$ we get exactly the same law for unrooted trees.
It remains to compute $\mu$ and $\sigma^2$. 

By using the procedure described in Appendix~\ref{sec-asymptotics} we get for our expample pattern
\[
\mu = \frac{5}{8e^3} = 0.0311169177 \dots
\]
and 
\[
\sigma^2= \frac{20e^3 + 72e^2+84e-175}{32e^6} = 0.0764585401\dots.
\]
We observe---as predicted by Theorem~\ref{Th1}---that both $\mu$ and~$\sigma^2$ can be written as
rational polynomials in $1/e$.

In what follows we will prove this fact (which completes the proof
of Theorem~\ref{Th1}) and also present an easy formula for $\mu$.
Unfortunately the procedure for calculating $\sigma^2$ is much
more complicated so that it seems that there is no simple formula.

\begin{proposition}\label{Pro4}
Let $x_0= 1/e$ and\/ $\aaa_0$ be given by Proposition~\ref{Pro3} and
let $P_j({\bf y},u)$ $(1\le j\le L)$ be the polynomials of
Proposition~\ref{Pro1}, with $\yy=(y_0,\dots,y_L)$. Then $\mu$ (of
Theorem~\ref{Th1}) is a polynomial in $1/e$ with rational coefficients and is given by
\begin{equation}\label{eqPro4}
\mu = \frac 1e \sum_{j=1}^{L} \frac{\partial P_j}{\partial u}
(\aaa_0,1).
\end{equation}
\end{proposition}

\begin{proof}
Let $\aaa = \FF(x,\aaa,u)$ be the system of functional equations of
Proposition~\ref{Pro1}. In Appendix~\ref{sec-asymptotics} the
following formula for the mean is derived:
\begin{equation}
\mu = \frac 1{x_0}\frac {\bbb^{\mathrm T}\FF_u(x_0,\aaa_0,
1)}{\bbb^{\mathrm T}\FF_x(x_0,\aaa_0,1)}.
\end{equation}
Here $\bbb^T$~ denotes a positive left eigenvector of
$\II - \FF_\aaa$, which is unique up to scaling.

From the equality
\[
\FF(x,\aaa,u) = \left(
\begin{array}{c}
x\left( e^{a_0+\cdots+a_L} - \sum_{j=1}^{L} P_j(\aaa,1)\right) \\
xP_1(\aaa,u) \\
xP_2(\aaa,u) \\
\vdots \\
xP_{L}(\aaa,u)
\end{array} \right),
\]
we get, after denoting $\frac{\partial P_i}{\partial a_j}$
with~$P_{i,a_j}$,
\begin{equation}\label{eqmatrix1}
\FF_{\aaa} = x \left(
\begin{array}{ccc}
e^{a_0+\cdots+a_L} - \sum_{j=1}^{L} P_{j,a_0} & \cdots &
e^{a_0+\cdots+a_L} - \sum_{j=1}^{L} P_{j,a_L} \\
P_{1,a_0} & \cdots & P_{1,a_L} \\
\vdots &  & \vdots  \\
P_{L,a_0} & \cdots & P_{L,a_L} \\
\end{array} \right).
\end{equation}
Since $a_0(x_0,1) + \cdots + a_L(x_0,1) = p(x_0,1) = 1$ we have
$x_0 e^{a_0(x_0,1) + \cdots a_L(x_0,1)} = 1$. Consequently the sum
of all rows of $\FF_\aaa$ equals $(1,1,\ldots, 1)$ for $x= x_0 =
1/e$. Thus, denoting the transpose of a vector~$v$ by~$v^{\mathrm
T}$, the vector $\bbb^{\mathrm T} = (1,1,\ldots, 1)$ is the unique
positive left eigenvector of $\II- \FF_\aaa$, up to scaling.

It is now easy to check that
\[
x_0 \bbb^{\mathrm T} \FF_x(x_0,\aaa_0,1) = \frac 1e e^{a_0(x_0,1) + \cdots
a_L(x_0,1)} = 1
\]
and that
\[
\bbb^{\mathrm T} \FF_u(x_0,\aaa_0,1) = \frac 1e \sum_{j=1}^{L}
P_{j,u}(\aaa_0,1).
\]
The fact that $\mu$ is a polynomial in $1/e$ is now a direct consequence 
from the fact that $\aaa_0$ consists of 
polynomials in $1/e$ and the fact that the coefficients are rational follows 
from the fact that $\FF(x,\aaa,u)$ has rational coefficients.
\end{proof}

Of course, with help of (\ref{eqPro4}) we can easily evaluate $\mu$
directly. As already indicated it seems that there is
no simple formula for $\sigma^2$.

Before proving Proposition~\ref{Pro5} we state in interesting fact
that will be used in the sequel.

\begin{lemma}\label{determinant}
Let $a_0,a_1,\ldots,a_L$ the partition of $p$ that is used in the proof 
of Theorem~\ref{Th1}. Then
\[
\det\left({\bf I} - {\bf F}_\aaa(x,\aaa,1)\right) = 1 - x e^{a_0+a_1+ \cdots+ a_L}.
\] 
\end{lemma}

Since the proof is a rather lengthy computation we postpone it to
Appendix~\ref{sec-determinant}.

\begin{proposition}\label{Pro5}
Let $x_0= 1/e$ and\/ $\aaa_0$ be given by Proposition~\ref{Pro3}. Then $\sigma^2$ (of
Theorem~\ref{Th1}) is a polynomial in $1/e$ (with rational coefficients).
\end{proposition}

\begin{proof}
From the proof of Proposition~\ref{Pro4} we already know that $x_u(1)$ can be represented
as a polynomial in $1/e$ (with rational coefficients). The next step is to show that
$\aaa_u(1)$ has the same property. For this purpose we have to look at the
system (\ref{eqyu}) 
\begin{align*}
(\II -\FF_\aaa)\aaa_u &= \FF_x x_u + \FF_u ,\\
-D_\aaa \aaa_u &=  D_x x_u + D_u,
\end{align*}
where $D(x,\aaa,u) = \det\left({\bf I} - {\bf F}_\aaa(x,\aaa,1)\right) = 
1 - x e^{a_0+a_1+ \cdots+ a_L}$. We first observe that
\[
D_\aaa(x_0,\aaa_0,1) = (-1, -1, \ldots, -1).
\]
Hence, we can replace the first row of the $(L+1)\times(L+1)$-matrix
$\II -\FF_\aaa$ (that is redundant since the
matrix has rank $L$)  by the row $(1,1,\ldots, 1)$ and obtain
a regular linear system for $\aaa_u(1)$. Note that all entries of
the right-hand side of this linear system can be represented as
polynomials in~$1/e$.

Let ${\bf M}(x,\aaa)$ denote the matrix obtained from $\II -\FF_\aaa(x,\aaa,1)$ by replacing
the first row by $(1,1,\ldots, 1)$. If follows from the 
proof of Lemma~\ref{determinant} that $\det {\bf M}(x,\aaa) = 1$.
Further all entries of ${\bf M}(x_0,\aaa_0)$ can be represented as
polynomials in $1/e$. Thus, ${\bf M}(x_0,\aaa_0)^{-1}$ has the same
property and consequently $\aaa_u(1)$ has this property, too.

From that it directly follows from (\ref{eqxuu}) that $x_{uu}$ is also
represented as a polynomial in $1/e$. (By definition, $b(x,\aaa,u)$ is
a rational polynomial of the entries of  $\II -\FF_\aaa$.)

With help of \eqref{eqsigma} this finally leads to a representaion 
of $\sigma^2$ as a polynomial in $1/e$.
\end{proof}

This finally completes the proof of Theorem~\ref{Th1}.

\medskip

\section{Extensions and Generalizations}\label{sec:extensions}

In what follows we list some obvious and some less obvious extensions
of our main result. For the sake of conciseness we do not present the
details.

\subsection{Several Patterns}
Let \Mk{}$_1$, $\ldots$, \Mk{}$_k$ be $k$ different patterns. Then the problem
is
to determine the joint (limiting) distribution of the number of occurrences of
\Mk{}${}_1$, $\ldots$, \Mk{}${}_k$ in trees of size~$n$. Using the same
techniques as above (introducing the forest of planted patterns
deduced from the patterns) we
again obtain a system of functional equations. The only difference is that
we now have to count occurrences of \Mk{}${}_1$, $\ldots$, \Mk{}${}_k$ with
different variables $u_1,\ldots, u_k$, which is done in the same
fashion as for a single $u$. In view of Theorem \ref{th:syseqns},
multiple variables~$u$ make no difference and we obtain a multivariate
Gaussian limiting distribution.

\subsection{Patterns Containing Paths of Unspecified Length}

It might also be interesting to consider patterns where specific
edges can be replaced by paths of arbitrary length. It turns out
that this case in particular is more involved since a  natural
partition of all planted rooted trees is now infinite.
Nevertheless it is possible to replace infinite series of such
classes by one new class and end up with a finite system.
Thus, this leads to a Gaussian limit law (as above).

\subsection{Filled and Empty Nodes}
In our model we have distinguished between internal (filled) and
external (empty) nodes of the pattern \Mk{}, where the degrees of
the internal (filled) nodes have to match exactly. It also seems to be
possible to consider the following more general matching problem:
Let \Mk{} again be a finite tree, where certain nodes are ``filled''
and the remaining ones are ``empty''. Now we say that \Mk{} matches
if it occurs as a subtree such that the corresponding degrees of the
filled nodes are equal whereas the degrees of the empty nodes might be
different. It seems that the counting procedure above can be adapted to cover
this case, too. However, it is definitely  more involved.
For example, if leaves of the pattern are filled nodes then these
nodes have to be leaves wherever the pattern occurs.
This implies that some of the functions $a_j(x,u)$ are then explicitly given
in the system and the dependency graph is not strongly connected.
However, it seems that this situation can be managed by eliminating these functions.
Furthermore, and this is more serious, in general one has
to consider infinitely many classes of trees leading to an
infinite system of functional equations, in particular if an internal node is ``empty''.
In such a case Theorem \ref{th:syseqns} cannot be applied
any more. Nevertheless we hope that the approach of Lalley \cite{lalley},
that is applicable to infinite systems of functional equations in one
variable, can be generalized to a corresponding generalization of
Theorem \ref{th:syseqns} to proper infinite systems. Thus, we
can expect a Gaussian limit law even in this case.

\begin{figure}[htb]
\centering
\includegraphics[width=3.6cm]{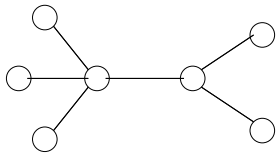}
\caption{Example pattern with empty nodes} \label{fig:emptynode}
\end{figure}

In order to be more precise we will present an easy example. Let $\Mk{}$
denote the pattern depicted in Figure~\ref{fig:emptynode}. Here all
nodes are empty. Thus, the corresponding pattern counting problem
is a subgraph counting problem.

We partition all planted trees according to their root degree.
Let $a_k$ denote the set of planted rooted trees with root out-degree $k$
and $a_k(x,u)$ the correponding generating function (that also counts
the number of subgraph occcurences of $\Mk{}$). Further, let $r(x,u)$
denote the generating function of rooted trees. Then we have
\[
a_k(x,u) = \frac x{k!} \left( \sum_{i\ge 0} a_i(x,u) 
u^{\binom k2 \binom i3 + \binom k3 \binom i2 } \right)^k \qquad (k\ge 0)
\]
and
\[
r(x,u) = x \sum_{k\ge 0} \frac 1{k!} \left( \sum_{i\ge 0} a_i(x,u) 
u^{\binom {k-1}2 \binom i3 + \binom {k-1}3 \binom i2} \right)^k.
\]
This system is easy to solve for $u=1$. Here we have
$a_k(x,1) = x p(x)^k/k!$ and $r(x,1) = p(x)$. By taking derivatives
with respect to $u$ and summing over all $k$ we also get (after some algebra)
\[
r_u(x,1) = \frac 5{12} \frac{p(x)^7}{1-p(x)} + \frac 16  \frac{p(x)^8}{1-p(x)} +
\frac {p(x)^7}6.
\]
This implies that the average value of pattern occurences (in this sense) is of the form 
$(7/12)n + O(1)$, that is, $\mu = 7/12$. In principle it is also possible to get asymptotics for
higher moments but the calculations get more and more involved.

\subsection{Simply Generated Trees}

Simply generated trees have been introduced by Meir and Moon \cite{MM}
and are proper generalizations of several types of rooted trees.
Let
\[
\varphi(x) = \varphi_0 + \varphi_1 x + \varphi_2 x^2 + \cdots
\]
be a power series with non-negative coefficients; in particular
we assume that $\varphi_0>0$ and $\varphi_j>0$ for some $j\ge 2$.
We then define the weight $\omega(T)$ of a finite rooted tree $T$
by
\[
\omega(T) = \prod_{j\ge 0} \varphi_j^{D_j(T)},
\]
where $D_j(T)$ denotes the number of nodes in $T$ with $j$ successors.
If we set
\[
y_n = \sum_{|T| = n} \omega(T)
\]
then the generating function
\[
y(x) = \sum_{n\ge 1} y_n x^n
\]
satisfies the functional equation
\[
y(x) = x \varphi(y(x)).
\]
In this context, $y_n$ denotes a weighted number of trees of size $n$.
For example, if $\varphi_j = 1$ for all $j\ge 0$ (that is, $\varphi(x) =
1/(1-x)$) then all rooted trees
have weight $\omega(T) = 1$ and $y_n = p_n$ is the number of planted plane
trees.
If $\varphi_j = 1/j!$ (that is, $\varphi(x) = e^x$)
then we formally get labeled rooted trees, etc.

Of course, we can proceed in the same way as above and obtain a
system of functional equations that counts occurrences of a specific
pattern in simply generated trees, and (under suitable conditions on
the growth of $\varphi_j$) we finally obtain a Gaussian limiting
distribution. This has explicitly been done by Kok in his thesis
 \cite{KOK05:aofa,KOK05:thesis}.

\subsection{Unlabeled Trees}

Let $\hat p_n$ denote the number of unlabeled planted rooted trees and $\hat
t_n$ the
number of unlabeled unrooted trees. The generating functions are denoted by
\[
\hat p(x)=\sum_{n\ge1}\hat p_n x^n\quad\mbox{and}\quad \hat t(x)=\sum_{n\ge1}
\hat t_n x^n.
\]
The structure of these trees is much more difficult than that of labeled trees.
It turns out that one has
to apply P\'olya's theory of counting and an amazing observation (\ref{eqTh62})
by
Otter \cite{Ot}.
The generating functions $\hat p(x)$ and $\hat t(x)$ satisfy the functional
equations
\[
\hat p(x) = x \sum_{k\ge 0} Z\left(S_k;\hat p(x),\hat p(x^2),\ldots, \hat
p(x^k)\right) \\
= x\exp\left(\hat p(x) + \frac 12 \hat p(x^2) + \frac 13 \hat p(x^3) + \cdots
\right)
\]
and
\begin{equation}\label{eqTh62}
\hat t(x)= \hat p(x)-\frac 12 \hat p(x)^2+\frac 12 \hat p(x^2),
\end{equation}
where $Z(S_k;x_1,\ldots,x_k)$ denotes the cycle index of the symmetric group
$S_k$.
These functions  have a common radius of convergence $\rho\approx0.338219$ and
a local expansion of the form
\[
\hat p(x)=1-b(\rho-x)^{1/2}+c(\rho-x)+d(\rho-x)^{3/2}+\Ord{(\rho-x)^2)}
\]
and
\[
\hat t(x)=\frac{1+\hat p(\rho^2)}2-\frac{b^2+2\rho \hat
p'(\rho^2)}2(\rho-x)+ bc(\rho-x)^{3/2}+\Ord{(\rho-x)^2)},
\]
where $b \approx 2.6811266$ and $c=b^2/3\approx2.3961466$, and
$x=\rho$ is the only singularity on the circle of convergence $|x|=\rho$.
Thus, they behave similarly as $p(x)$ and $t(x)$.
We also get
\[
\hat p_n =\frac{b\sqrt\rho}{2\sqrt\pi}n^{-3/2}\rho^{-n}\left( 1 +
\Ord{n^{-1}}\right)
\]
and
\[
\hat t_n=\frac{b^3\rho^{3/2}}{4\sqrt\pi}n^{-5/2}\rho^{-n}\left( 1 +
\Ord{n^{-1}}\right).
\]

Furthermore, it is possible to count the number of nodes of specific
degree with the help of bivariate generating functions (compare with
\cite{DrGi99}). Thus, using P\'olya's theory of counting we can also
obtain a system of functional equations for bivariate generating
functions that count the number of occurrences of a specific
pattern. The major difference to the procedure above is that this
system also contains terms of the form $a_j(x^k,u^k)$ for $k\ge 2$.
Fortunately these terms can be considered as  known functions when
$x$~varies around the singularity $\rho$ and $u$ varies around~$1$
(compare again with \cite{DrGi99}). Hence, Theorem~\ref{th:syseqns}
applies again and we can proceed as above. This has explicitly been
done by Kok in his thesis  \cite{KOK05:aofa,KOK05:thesis}.

\subsection{Forests}

First, let us consider the case of labeled trees with generating function
$t(x,u)$. Then the generating function $f(x,u)$ of unlabeled forests is
given by
\[
f(x,u) = e^{t(x,u)}.
\]
Thus, the singular behaviour of $f(x,u)$ is the same as that of $t(x,u)$
(compare with \cite{DrGi99}) and consequently we again obtain a
Gaussian limiting distribution for the number of occurrences of a specific
pattern in labeled forests.

The case of unlabeled forests is similar. Here we have
\[
\hat f(x,u) =  \exp\left(\hat t(x,u) + \frac 12
\hat t(x^2,u^2)+ \frac 13 \hat t(x^3,u^3) + \cdots \right).
\]

Of course, we can consider other classes of trees or forests
of a given number of trees.

\subsection{Forbidden Patterns}

It is also interesting to count the number $t_{n,0}$ of trees
of size $n$ without
a given pattern. The generating function of these numbers is
just $p(x,0)$, resp.\ $t(x,0)$. It is now an easy exercise to
show that there exists an $\eta>0$ such that
\[
t_{n,0} \le t_n e^{-\eta n}.
\]
The only thing we have to check is that the radius of convergence
of $t(x,0)$ is larger than the radius of convergence of $t(x,1)$.
However, this is obvious since the radius of convergence of
$t(x,u)$ (which is the same as that of $p(x,u)$) is given by
$x(u)$ (for $u$ around $1$) and $x'(1)<0$.

\begin{appendix}

\section{Algorithms}\label{sec-algorithms}

\begin{figure}[htb]
\centering
\includegraphics[width=6cm]{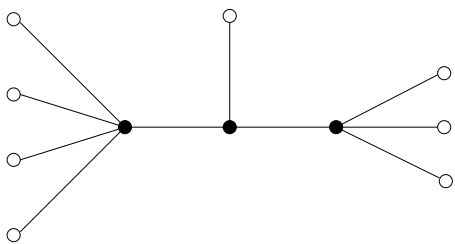}
\caption{Example pattern $\mathcal M$} \label{fig:apppattern}
\end{figure}

In the main part of this paper we showed that the limiting
distribution of the number of pattern occurrences is normal with
computable $\mu$ and $\sigma^2$. However the family of classes
$\{a_0,a_1,\dots,a_L\}$ considered in the first part was especially
created to make the arguments more transparent, there were no
considerations about minimality. In this appendix we focus on
creating another partition $\mathcal A = \{a_0,\dots,a_L\}$ of
$p$ which has considerably less classes. It also has the properties that it
is recursively describable and allows an
unambiguous definition of the number of additional occurrences
$K(l_0,\dots, l_L)$ of the pattern.
For example we show that for the pattern of Figure~\ref{fig:apppattern} we need 
just 8 equations whereas the previous proof would use more than 1000 equations.

First we remark that in some
cases it is profitable to adjust the structure of the system
of equations \eqref{eq:asystem} in Proposition \ref{Pro1} by
allowing an additional polynomial $P_0(y_0,\dots,y_L,u)$ in the
first equation. The first equation then becomes
\begin{equation*}
\begin{aligned}
a_0(x,u) &=  x\cdot P_0(a_0(x,u),\ldots,a_L(x,u),u)  \\
&\qquad+(xe^{a_0(x,u)+ \cdots + a_L(x,u)} - x \sum_{j= 0}^{L}
P_j(a_0(x,u),\ldots,a_L(x,u),1)).
\end{aligned}
\end{equation*}
This system still fits our analytical framework. The advantage is
that for example the \emph{minimal} system of equations
for counting stars in trees on  page \pageref{eq:startrees} now fits
this
modified system. 

\medskip

The idea for constructing $\mathcal A$ will be to create in a first
time a certain family of tree classes $\mathcal S= \{
t_1,\dots,t_n\}$, not necessarily building a partition of $p$. Each
of these classes will be defined as the class of all trees in $p$
which ``start'' in a certain way, or with other words, which match a
certain tree $t_i'$ at the root, just as was the case for the $a_i$
in the main part of this paper. By abuse of notation we will usually
write $t_i$ instead of $t_i'$ for this tree. Let $J=\{1,\dots,n\}$
and $t_i^c = p \setminus t_i$. Now, by collecting in $\mathcal A$
all different, non-empty classes of the form
\begin{align}\label{aidef}
a_I = \bigcap_{i\in I} t_i \cap \bigcap_{i\in J\setminus I}
t_i^c,\qquad I\subseteq J
\end{align}
we will obtain a partition $\mathcal A$ of $p$. This partition will
have a recursive description by construction, see the algorithms
below. Furthermore, if $\mathcal S$ is sufficiently rich, this
partition will allow an unambiguous definition of $K(l_0,\dots,l_L)$.

\medskip

We now make some considerations about the properties that
$\mathcal S$ should possess to make sure that $\mathcal A$ will
allow an unambiguous definition of $K(l_0,\dots,l_L)$.  Let $b$ be a
subclass of $p$. For each tree $T\in p$ we can determine the number
$k(T)$ of pattern occurrences at the root of $T$. Let
$k({b})=\{\,k(T):T \in b\,\}$. Because the patterns have finitely
many nodes
and because in each internal node the degree is fixed and the root
has to be part of the match, there are only finitely many ways for a
pattern match. Thus the set $k({b})$ will be finite and non-empty.
Now let $a_I$ defined by equation \eqref{aidef} (and non-empty). Now it holds that
\begin{align}\label{kai}
k(a_I) \subseteq \bigcap_{i\in I} k(t_i) \cap \bigcap_{i\in
J\setminus I} k(t_i^c)
\end{align}
because a tree $T$ in $a_I$ is by definition in $t_i,\ i\in I$ and
$t_i^c,\ i\in J\setminus I$, thus the number of pattern occurrences
at the root is constrained by $k(t_i),\ i\in I$ and $k(t_i^c),\ i\in
J \setminus I$. If $\mathcal S = \{ t_1,\dots, t_n\}$ is
sufficiently rich, then $k(a_I)$ will only consist of a single
number. This will be the case if for each $m\in\mathbb N$, the
family $\mathcal S$ contains
all classes of trees ``starting'' with all possible arrangements of $m$
overlapping patterns. Indeed, if we have for example for a certain
tree class $t_i$ that $k(t_i) =\{ r,r+1\}$, then there will be
another tree class $t_j$, which is a subclass of $t_i$ with
$k(t_j)=\{ r+1 \}$. Now the intersections $b=t_i \cap t_j^c$ and $c=t_i \cap t_j$
 will yield tree classes with a singleton $k(.)$,
namely $k(b)= \{r\}$ and $k(c)=\{r+1\}$.

\begin{figure}[htb]
\centering
\includegraphics[width=10cm]{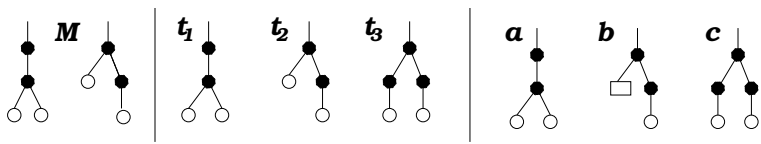}
\caption{On the left: Planted patterns. Center: Classes $t_i$. Right: Classes 
$\{a,b,c\}$. The white box here means a node of out-degree different from~1. Note: 
this does not correspond to the output of the algorithms of this appendix}
\label{fig:kex}
\end{figure}

 For example consider a pattern which consists of a node of degree~2 
attached to a node of degree~3. The corresponding planted patterns are shown 
in Figure~\ref{fig:kex}. Now let  $\mathcal S$ consist of the three classes 
$t_1,t_2,t_3$, shown in the center of Figure~\ref{fig:kex}. We have $k(t_1)=\{1\}$, because the left planted pattern surely matches and the other does not, $k(t_2)=\{1,2\}$, because the left planted pattern does not match and the right one matches at least once, but possibly twice. $k(t_3)= \{2\}$, because the left pattern does not match and the right one surely matches twice. We see that the only non-empty intersections of the form \eqref{aidef} are $a=t_1 \cap t_2^c \cap t_3^c$, $b=t_1^c\cap t_2 \cap t_3^c$ and $c=t_1^c\cap t_2 \cap t_3$. We obtain $k(a)=k(b)=\{1\}$ and $k(c)=\{2\}$, which are all singletons.
Because we also need a recursive description of the final partition $\mathcal A$, 
we will construct some additional tree classes $t_i$. As the partition becomes 
finer when dealing with more classes $t_i$, it is clear that $k$ remains well-defined. 

On the other hand we do not have to associate a unique number to $k(a_I)$,
only to $K(l_0,\dots,l_L)$. Therefore we can slightly reduce the family
$\mathcal S= \{t_1,\dots,t_n\}$. In the algorithm below this reduction
of $\mathcal S$ corresponds to considering only proper subtrees of the trees 
$q\in\mathcal Q$ ($q$ itself is excluded).

\medskip

A coarse-grain description of an algorithm now follows.

\begin{enumerate}
\item \label{alg:planemb} Calculate the set $\mathcal U$ of all planar embeddings of all planted patterns
deducible from the pattern $\mathcal M$.

\item\label{step2}\label{alg:construct} Consider the planted planar
trees issue of step \ref{alg:planemb} as planar tree classes and
take all possible intersections of any number of those classes. Now take the implied non-planar general tree structure of each class and collect these non-planar planted trees in the set $\mathcal Q$.

\item \label{alg:dag} Create a family $\mathcal S=\{t_1,\dots,t_n\}$ for the
forest of planted subtrees of trees  $q\in\mathcal Q$, excluding the trees $q$
themselves, where each $t_j$ has a recursive description in $t_0, t_1,\dots,
t_{j-1}$ and where $t_0$ denotes a leaf.

\item \label{alg:disamb} Now interpret $t_0$ as the class of all trees $p$ and interpret the trees $t_i\in\mathcal S$ as non-planar tree classes. Construct a partition $\mathcal A= \{ a_0,\dots,a_L \}$ of the class of all planted trees $p$ together with a recursive description (compare with \eqref{aidef}).

\item \label{alg:k} Calculate for each term in the recursive description the
number $K(l_0,\dots,l_L)$ of additional pattern occurrences and
deduce a system of equations for the generating functions $a_j(x,u)$
of the classes $a_j$.

\end{enumerate}

Before giving more detailed algorithms, we give an example.
Consider the pattern of Figure~\ref{fig:apppattern}.

With the procedure of the main part of the article we would end up
with more than 1000 classes, yielding a system of equations with the
same number of equations. However, by using the following refined algorithm we only need 8
classes.

\begin{figure}[htb]
\centering
\includegraphics[width=10cm]{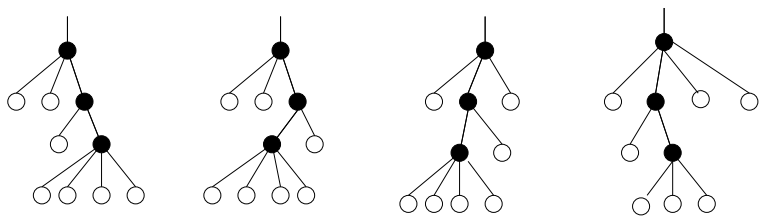}
\caption{Some of in total 16 planted planar embeddings $\mathcal U$} \label{fig:planemb}
\end{figure}

In the first step we create all planar embeddings of the corresponding planted pattern (trees $\tau_1,\tau_2,\tau_3$ of Figure~\ref{fig:kcalc}). This yields $3\cdot 2 + 2 + 4\cdot 2 = 16$ planar
trees of which some are shown in Figure~\ref{fig:planemb}.

\begin{figure}[htb]
\centering
\includegraphics[width=10cm]{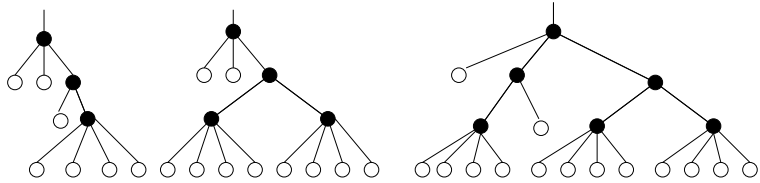}
\caption{Some of in total 24 non-planar trees of $\mathcal Q$}
\label{fig:intersec}
\end{figure}

We now consider these structures as planar tree classes and
additionally construct tree classes by taking all possible
intersections of any number of the classes issued from step~1. Then, we take the 
non-planar implied tree structure of each planar class and collect these 
trees in $\mathcal Q$. We end up with 24
different trees: 9~that stem from~$\tau_1$, 1~from~$\tau_2$, and 14~from~$\tau_3$. Some of
them are shown in Figure~\ref{fig:intersec}.

\begin{figure}[htb]
\centering
\includegraphics[width=10cm]{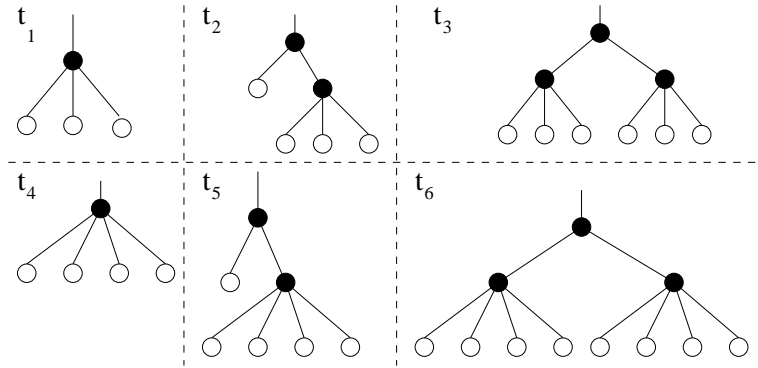}
\caption{Non-planar trees $t_i$ which possess a recursive description} \label{fig:dag}
\end{figure}

For all proper subtrees for each tree in $\mathcal Q$ we now construct a recursive description. 
For example, for the leftmost tree of Figure~\ref{fig:intersec} we first consider the subtree consisting of a node with four 
leaves. We denote this class by $t_4=xt_0^4$. (Here we use the following structural
notation: $x$ denotes a root node, $t_0$ a leaf and $xt_0^4$ denotes a root to which are attached
4 leaves.) The next subtree is a root of out-degree 2 to which a subtree of type $t_4$ is attached.
We denote this with $t_5=xt_0t_4$. Figure~\ref{fig:dag} shows all 6 trees we end up with.  Observe on our example that the collection of subtrees at the root extracted from the 24~trees in~$\mathcal Q$ consists of only 6~trees.

Their recursive descriptions are given by

\begin{gather}\label{eq:dag}
t_1 = xt_0^3,\quad t_2 = xt_0 t_1, \quad t_3=xt_1^2,\quad
t_4=xt_0^4,\quad t_5=xt_0t_4,\quad t_6=xt_4^2.
\end{gather}

\begin{figure}[htb]
\centering
\includegraphics[width=10cm]{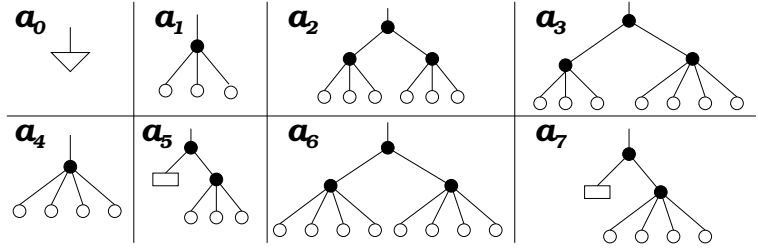}
\caption{Non-planar partition classes. The white box means ``not out-degree 3 or~4'' and the white triangle means ``anything that is not contained in
the other classes''} \label{fig:part}
\end{figure}

We now interpret $t_0$ in \eqref{eq:dag} as the class of all planted trees $p$. 
The other $t_i$ are also interpreted as tree classes. For example, $t_1$ is 
the class of all trees with root out-degree 3.
We now construct a partition based on these classes and their recursive
description of \eqref{eq:dag}. We obtain the classes of Figure
\ref{fig:part}.

Their recursive description is given by
\begin{equation}\label{eq:part}
\begin{aligned}
a_0 &=p\setminus \bigoplus_{i=1}^7 a_i =x\oplus x\bigoplus_{i=0}^{7} a_i 
\oplus x(a_0\oplus a_2\oplus a_3\oplus a_5\oplus a_6 \oplus a_7)^2 \oplus 
x\bigoplus_{n=5}^\infty \left( \bigoplus_{i=0}^{7} a_i \right)^n, \\
a_1 &=xp^3, \\
a_2 &=xa_1^2, \\
a_3 &=xa_1a_4, \\
a_4 &=xp^4, \\
a_5 &=x(a_0\oplus a_2\oplus a_3\oplus a_5\oplus a_6 \oplus a_7)a_1, \\
a_6 &=xa_4^2, \\
a_7 &=x(a_0\oplus a_2\oplus a_3\oplus a_5\oplus a_6 \oplus a_7)a_4.
\end{aligned}
\end{equation}

The last step consists of determining the number of additional
occurrences $K(l_0,\dots,l_7)$ for each term in the recursive
description \eqref{eq:part} and translating \eqref{eq:part} in a
system of equations for the generating functions $a_j(x,u)= a_j$.
As an example we consider the equation for $a_1$. Class $a_1$ consists
of the trees of root out-degree 3. We get no additional occurrences of the pattern if
we attach a tree of class $a_0,a_1,a_2,a_4$ or $a_5$ to such a root, we get
one additional occurrence for each tree of class $a_3$ or $a_7$ and we have two
additional occurrences for each tree of class $a_6$ attached to the root. This yields the equation
for $a_1(x,u)$ below. Altogether we obtain:
\begin{align*}
a_0 &= x+ x\sum_{i=0}^7 a_i + \frac12 x (a_0 + a_2 + a_3 + a_5 + a_6
+ a_7)^2 + x\sum_{n\ge 5} \frac1{n!} \left(\sum_{i=0}^7 a_i\right)^n,
\\
a_1 
    &= \frac1{3!}x(a_0+a_1+a_2+a_4+a_5 + (a_3+a_7)u +a_6u^2 )^3, \\
a_2 &= \frac12 xa_1^2,\\
a_3 &= xa_1a_4u,\\
a_4 
    &= \frac 1{4!}x(a_0+a_1+a_4+a_6+a_7 + (a_3+a_5)u + a_6u^2)^4, \\
a_5 &= x(a_0 + a_2 + a_3 + a_5 + a_6 +  a_7)a_1, \\
a_6 &= \frac12 xa_4^2,  \\
a_7 &= x(a_0 + a_2 + a_3 + a_5 + a_6 +  a_7)a_4.
\end{align*}
We can now calculate $\mu$. We get 
$\mu = \frac{256-43e}{8e^3} = 0.865759040\dots$. The computation of 
$\sigma^2$ was not feasible, because of memory 
problems.\footnote{The actual computation uses polynomial expressions with more than 200,000 terms. We used Maple 9.5, which used up the memory of 1 GB and a very large part of the 1 GB swap.}

\subsection{Planar embedding algorithm: GeneralToPlanar} 
\label{app:A1}
$ $\\
\noindent\emph{Input:\/} a general planted tree $\tau$

\noindent\emph{Output:\/} the set $\mathcal U$ of planted planar
trees~$\pi$ that share~$\tau$ as their implied general tree
structure

\noindent\emph{Algorithm:}
\begin{enumerate}

\item write~$\tau$ in the form~$\nc\tau_1\dotsm\tau_k$, that is, let
$k$ be the root out-degree of~$\tau$ and $\tau_1$,~\dots, $\tau_k$
be the children at the root

\item for each~$i$ between 1 and~$k$, recursively compute
$P_i=\mathrm{GeneralToPlanar}(\tau_i)$

\item construct and return the set of planar trees
$\nc\pi_{\sigma(1)}\dotsm\pi_{\sigma(k)}$ over all choices
of~$\pi_i\in P_i$ and over all permutations~$\sigma$
of~$\{1,\dots,k\}$

\end{enumerate}

\subsection{Tree class intersection algorithm}
$ $\\
\noindent\emph{Input:\/} a set of planted planar trees $\mathcal U$

\noindent\emph{Output:\/} the set $\mathcal
Q$ of non-planar planted trees which are obtained by intersecting planar
tree classes based on $\mathcal U$ and collecting the non-planar tree 
structures of the resulting planar tree classes.

\noindent\emph{Algorithm:}
\begin{enumerate}

\item For each $i$ between 1 and $|\mathcal U|$, consider all
$i$-tuples of different trees $\pi_1,\dots,\pi_i \in \mathcal U$ and
determine for each $i$-tuple if $s=\pi_i\cap \dots \cap \pi_i$ may
be interpreted as a non-empty tree class. In that case, let $s'$ be
the implied non-planar tree structure of $s$ and add $s'$ to the set
$\mathcal Q$.

\end{enumerate}

\subsection{DAGification algorithm}
$\ $\\
We construct a recursive description for the forest of planted
subtrees for each tree in a given set of planted trees. Here we
do not consider the tree itself as a subtree of itself. This calculation
is reminiscent of the DAGification process of computer science (see,
e.g., \cite{ASU86}), which aims at compacting an expression tree by
sharing repeated subexpressions. However, if we interpret those
subtrees as classes, the intersection of two classes need not be
empty.

\bigskip

\noindent\emph{Input}: set of planted trees $\mathcal Q$

\bigskip

\noindent\emph{Output}: a number $m$ and a recursive description of
the forest of planted subtrees $\mathcal S=\{ t_1,\dots, t_m\}$ of 
the trees of $\mathcal Q$, of the
form
\begin{equation*}
t_i=\nc t_{\lambda_1^{(i)}}\dotsm t_{\lambda^{(i)}_{r_i}} \qquad(r_i
\in\mathbb N) \qquad\text{for~$1\leq i\leq m$}
\end{equation*}
with the constraint $\lambda_j^{(i)}<i$ for all $i$ and~$j$

\bigskip

\noindent\emph{Algorithm}:

\begin{list}{}{}

\item[(Initialization)] Introduce the exceptional type~$t_0$ to
denote the planted tree consisting of a single node (in other words,
a leaf) and set $m$ to $1$

\item[(Main loop)] For all planted trees of $\mathcal U$
perform a depth-first traversal of the tree, starting from the
planted root; during this recursive calculation, at each
node~$n$:

\begin{enumerate}

\item if the node is a leaf, return the type~$t_{0}$

\item else, recursively determine the type associated with each child
of~$n$

\item If $n$ is a not the planted root of the tree,
write the subtree rooted at~$n$ as a (commutative) product $\pi=\nc
t_{\lambda_1}\dotsm t_{\lambda_r}$ of the types obtained in the
previous step

\item look up the uniquification table to check whether this product
has already been assigned a type~$t_{i}$

\item if not existent, increment $m$, create a new type~$t_{m}$, remember its
definition~$t_{m}=\pi$, and assign~$t_{m}$ to the product~$\pi$ in
the uniquification table.

\item return the type~$t_{i}$ if it was found by lookup, otherwise return
$t_{m}$

\end{enumerate}

\item[(Conclusion)] Return $m$ and the sequence of definitions of the
form~$t_{i}=\pi$, for~$i=1,2,\dots,m$.

\end{list}

\subsection{Disambiguating algorithm}
$\ $\\
The idea of the algorithm below is to consider each class of trees,
$t_i$, in turn, introducing its defining equation
\begin{equation*}
t_i=\nc t_{\lambda_1^{(i)}}\dotsm t_{\lambda_{r_i}^{(i)}}
\qquad(r\in\mathbb N)
\end{equation*}
into the calculation, while maintaining (and refining) a partition
\[p=a_0\oplus\dots\oplus a_L\]
of the total class of planted trees.  To be able to do so, it is
crucial that the recursive equation for~$t_i$ refers to
classes~$t_j$ with~$j<i$ only, starting with the special
class~$t_0=p$, the full class of planted trees.

At any stage in the algorithm, the class of $r$-ary trees is given
as the disjoint union of Cartesian products
\begin{equation*}
\bigoplus_{\lambda\in\Lambda}\nc t_{\lambda_1}\dotsm t_{\lambda_r}
\qquad\text{where}\qquad \Lambda=\{\,\lambda:\ell(\lambda)=r,\ 0\leq
\lambda_j\leq L\,\},
\end{equation*}
where $\ell(\lambda)$ denotes the number of components in the tuple
$\lambda$. In the process of the algorithm below, each class~$t_i$
gets represented in a ``polynomial'' form like above, summed over a
subset~$\Lambda$ of the set of integer sequences $\lambda =
(\lambda_1,\dots, \lambda_r)$ of a given length~$r$. Computing
intersections and differences of classes means merely computing
intersections and differences of the~$\Lambda$ in their
representations, because of the recursive structure of the input and
of the algorithm itself.

\bigskip

\noindent\emph{Input}:

\begin{itemize}

\item A family $\mathcal S=\{ t_1,\dots,t_m\}$ of classes of trees with recursive descriptions of the form
\begin{equation*}
t_i=\nc t_{\lambda_1^{(i)}}\dotsm
t_{\lambda_r^{(i)}}\qquad(r=\ell(\lambda^{(i)}))
\qquad\text{for~$1\leq i\leq m$}
\end{equation*}
with the constraint $\lambda_j^{(i)}<i$ for all $i$ and~$j$

\end{itemize}

\bigskip

\noindent\emph{Output}:

\begin{itemize}

\item an integer~$L$ implying a partition
\[p=a_0\oplus\dots\oplus a_L\]

\item a representation of each~$t_i$ of the form
\begin{equation*}
t_i=\bigoplus_{j\in I_i}a_j \qquad\text{for $0\leq i\leq m$
and~$I_i\subseteq\{0,\dots,L\}$}
\end{equation*}

\item a recursive description of the~$a_i$ of the form
\begin{equation*}
a_i=\bigoplus_{\lambda\in\Lambda_i} \nc a_{\lambda_1}\dotsm
a_{\lambda_{\ell(\lambda)}} \qquad\text{for~$1\leq i\leq L$},
\end{equation*}
$a_0$~being implicitly described as~$p\setminus(a_1\oplus\dots\oplus
a_L)$

\end{itemize}

\bigskip

\noindent\emph{Algorithm}:

\begin{list}{}{}

\item[(Initialization)] Start with the trivial partition $p=a_0$
for~$L=0$, the single representation $t_0=a_0$, that is,
$I_0=\{0\}$.

\item[(Main loop)] For~$k$ from~1 to~$m$ do

\begin{enumerate}

\item replace each~$t_i$ in the definition of~$t_k$ with its current
representation in terms of the~$a_j$, expand, and set~$s$ to the
result, so as to get a representation of~$t_k$ of the form
\begin{equation*}
s=\bigoplus_{\lambda\in\Lambda^{(s)}} \nc a_{\lambda_1}\dotsm
a_{\lambda_{\ell(\lambda)}} \qquad\text{for some~$\Lambda^{(s)}$}
\end{equation*}

\item for~$i$ from~1 to~$L$ while~$s\neq\emptyset$ do

    \begin{enumerate}

    \item set~$b$ to~$a_i\cap s$ by setting $\Lambda_\cap$
    to~$\Lambda_i\cap\Lambda^{(s)}$

    \item\label{non-empty-intersection} if $b\neq\emptyset$, then do

    \begin{enumerate}

    \item set~$b'$ to~$a_i\setminus s$

    \item if $b'\neq\emptyset$, then

        \begin{enumerate}

        \item create a new $a_j$ with description~$b'$:
        increment~$n$ before setting~$a_L$ to~$b'$, that is,
        before setting~$\Lambda_L$
        to~$\Lambda_i\setminus\Lambda^{(s)}$

        \item split~$a_i$ into $a_i\oplus a_L$ in the representations
        of the~$t_j$, that is, add~$n$ into each set~$I_j$
        containing~$i$

        \item split $a_i$ into $a_i\oplus a_L$ in the
        descriptions of the~$a_j$, $b$, and~$s$, that is, for
        each sequence in each of the~$\Lambda_j$, $\Lambda_\cap$,
        and~$\Lambda^{(s)}$, add sequences with $i$ replaced by
        $L$ when the sequence involves~$i$ (if $i$ occurs more than
        once, then replace $i$ by $i$ or $L$ in all possible ways)

        \item set~$a_i$ to~$b$ by setting~$\Lambda_i$
        to~$\Lambda_\cap$

        \end{enumerate}

    \item set~$s$ to~$s\setminus b$, which is also $s\setminus
    a_i$, and update~$\Lambda^{(s)}$ by setting it
    to~$\Lambda^{(s)}\setminus\Lambda_i$

    \end{enumerate}

    \end{enumerate}

\item if $s\neq\emptyset$, then

    \begin{enumerate}

    \item\label{create-remainder} create a new~$a_j$ with
    description~$s$: increment~$L$ before setting~$a_L$ to~$s$, that
    is, before setting~$\Lambda_L$ to~$\Lambda^{(s)}$

    \item split~$a_0$ into $a_0\oplus a_L$ in the representations
    of the~$t_j$, that is, add~$L$ into each set~$I_j$
    containing~$0$

    \item split $a_0$ into $a_0\oplus a_L$ in the descriptions of
    the~$a_j$, that is, for each sequence in each of the~$\Lambda_j$,
    add sequences with $0$ replaced by
    $n$ when the sequence involves~$0$ (if $0$ occurs more than
    once, then replace $0$ by $0$ or $L$ in all possible ways)

    \end{enumerate}

\item represent~$t_k$ as the union of all those~$a_i$s that have
contributed a non-empty~$b$ at step~(\ref{non-empty-intersection})
and of~$a_L$ if a new~$a_j$ was created at
step~(\ref{create-remainder}), that is, create the corresponding
set~$I_k$ consisting of the contributing~$i$s, together with~$L$ if
relevant

\end{enumerate}

\item[(Final step)] Return~$L$, the representations of the~$t_i$
for~$1\leq i\leq m$, the recursive descriptions of the~$a_i$
for~$1\leq i\leq L$

\end{list}

We will explicitly show the stages through which the algorithm goes
when running with the input $\eqref{eq:dag}$. For readability, we will
keep expressions in factored form.
\begin{list}{}{}

\item[$k=1$:] from $t_1=xa_0^3$, we derive $t_1=a_1$ and
$a_1=x (a_0\oplus a_1)^3$.

\medskip

\item[$k=2$:] from $t_2=x(a_0\oplus a_1)a_1$, we derive
$t_1=a_1,\ t_2=a_2$ and $a_1=xp^3,\ a_2=xpa_1$, where $p=a_0\oplus
a_1\oplus a_2$.

\medskip

\item[$k=3$:] from $t_3=xa_1^2$, we derive $t_1=a_1,\ t_2=a_2\oplus
a_3,\ t_3=a_2$ and $a_1=xp^3,\ a_2=xa_1^2,\ a_3=x(a_0\oplus
a_2\oplus a_3)a_1$, where $p=a_0\oplus a_1\oplus a_2\oplus a_3$.

\medskip

\item[$k=4$:] from $t_4=x(a_0\oplus a_1\oplus a_2\oplus a_3)^4$, we
derive
$t_1=a_1,\ t_2=a_2\oplus a_3,\ t_3=a_2,\ t_4=a_4$ and $a_1=xp^3,\
a_2=xa_1^2,\ a_3=x(a_0\oplus a_2\oplus a_3 \oplus a_4)a_1,\
a_4=xp^4$, where $p=a_0\oplus a_1\oplus a_2\oplus a_3\oplus a_4$.

\medskip

\item[$k=5$:] from $t_5=x(a_0\oplus a_1\oplus a_2\oplus a_3\oplus
a_4)a_4$, we derive $t_1=a_1,\ t_2=a_2\oplus a_3\oplus a_5,\
t_3=a_2,\ t_4= a_4,\ t_5=a_3\oplus a_6$ and $a_1=xp^3,\ a_2=xa_1^2,\
a_3=xa_1a_4,\ a_4=xp^4,\ a_5=x(a_0\oplus a_2\oplus a_3\oplus
a_5\oplus a_6)a_1,\ a_6=x(a_0\oplus a_2\oplus a_3\oplus a_4\oplus
a_5\oplus a_6)a_4$, where $p=a_0\oplus a_1\oplus a_2\oplus a_3\oplus
a_4\oplus a_5\oplus a_6$.

\medskip

\item[$k=6$:] from $t_6=xa_4^2$, we derive $t_1=a_1,\ t_2=a_2\oplus
a_3\oplus a_5,\
t_3=a_2,\ t_4= a_4,\ t_5=a_3\oplus a_6 \oplus a_7,\ t_6=a_6$ and
$a_1=xp^3,\ a_2=xa_1^2,\ a_3=xa_1a_4,\ a_4=xp^4,\ a_5=x(a_0\oplus
a_2\oplus a_3\oplus a_5\oplus a_6 \oplus a_7)a_1,\ a_6= xa_4^2,\
a_7=x(a_0\oplus a_2\oplus a_3\oplus a_5\oplus a_6 \oplus a_7)a_4$,
where $p=a_0\oplus a_1\oplus a_2\oplus a_3\oplus a_4\oplus a_5\oplus
a_6\oplus a_7$.

\end{list}

\subsection{Calculation of $K(l_0,\dots,l_L)$: CountRootOccurrences}
$ $\\
\noindent\emph{Input:\/} non-planar planted trees $\tau$,
$\tau_1$,~\dots, $\tau_k$

\noindent\emph{Output:\/} the number of occurrences of any of
the~$\tau_i$ at the root of~$\tau$

\noindent\emph{Algorithm:\/}
\begin{enumerate}

\item fix one element~$\pi'$ from $\mathrm{GeneralToPlanar}(\tau)$  (see algorithm
\ref{app:A1})

\item for each~$i$ between 1 and~$k$, compute $P_i=\mathrm{GeneralToPlanar}(\tau)$

\item count and return the number of pairs~$(\pi_i,\pi')$ such that
$\pi_i$~is element of~$P_i$ and $\pi_i$~occurs at the root of~$\pi'$

\end{enumerate}

\begin{figure}[htb]
\centering
\includegraphics[width=10cm]{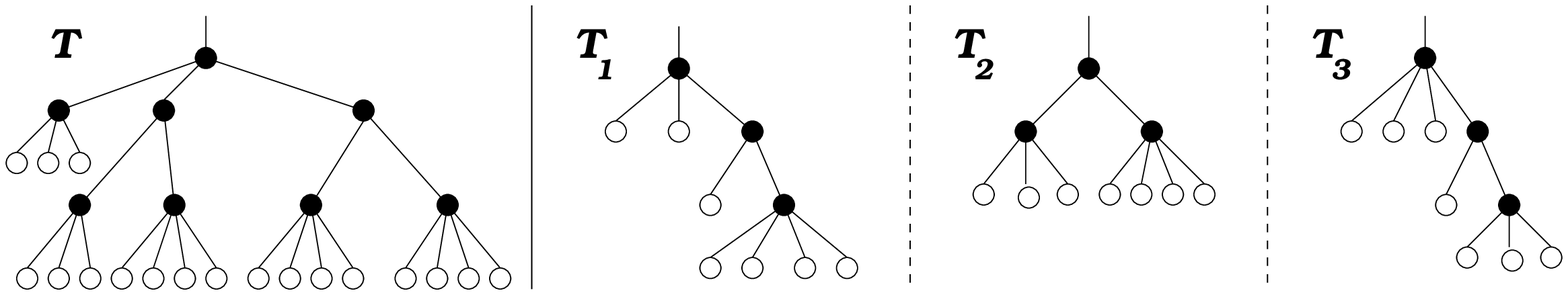}
\caption{Input trees $\tau, \tau_1, \tau_2, \tau_3 $} \label{fig:kcalc}
\end{figure}

As an example we calculate $K(0,1,0,1,0,0,1,0)$. This corresponds to calculating the 
number of additional occurrences in the class $xa_1a_3a_6$. The input trees $\tau,\tau_1,\tau_2,\tau_3$
are shown in Figure~\ref{fig:kcalc}. Here $\tau$ corresponds to the class  $xa_1a_3a_6$ 
and $\tau_1,\tau_2,\tau_3$ correspond to the three possible ways of planting the example pattern.

We take as fixed planar embedding $\pi'$ of $\tau$  the embedding of Figure
\ref{fig:kcalc}. We now iterate over the different
planar embeddings $\pi_1$ of~$\tau_1$ (6~of them), $\pi_2$ of~$\tau_2$ (2~of them), and $\pi_3$ of~$\tau_3$ (8~of them), and determine for each~$\pi_i$ ($i\in \{1,2,3\}$) whether it occurs at the root
of~$\pi'$. Consider for example the four embeddings shown in Figure~\ref{fig:planemb} (three embeddings of $\tau_1$, one embedding of $\tau_3$). The leftmost 
embedding matches $\pi'$, the one next to it as well. The third one does not match~$\pi'$, because the node with
out-degree four is in the wrong position. The rightmost embedding clearly does not match either. By considering all 
embeddings and counting the matches we get $k=K(0,1,0,1,0,0,1,0)= 3$.

The algorithm calculates the correct value of $k$, because the partition consisting of the
classes~$a_i$ is sufficiently fine. From this follows that every match above of a planar embedding really
gives rise to exactly one additional pattern occurrence. See the considerations made at the beginning of this
appendix.

By now the transformation to a systems of equations is easy. We get
the terms by replacing a term $x a_{j_1} \dotsm a_{j_s}$ in the
recursive description of $a_j$ by a term $xy_{j_1}\dotsm
y_{j_s}u^{K(l_0,\dots,l_L)}/l_0!\dots l_L!$. Here it is assumed that
terms that represent the same tree classes (like $xa_1a_2$ and
$xa_2a_1$) are identified before. It is clear that there are
only finitely many terms for which $K(l_0,\dots,l_L)$ might be
non-zero a priori.

\section{Asymptotics of Analytic Systems}\label{sec-asymptotics}

The following theorem is a slightly modified version of the main
theorem from \cite{Dr97}.  We denote the transpose of a vector~$v$
by~$v^{\mathrm T}$.  Let $\FF(x,\yy,\uu) =
(a_{10}(x,\yy,\uu),\ldots,F_N(x,\yy,\uu))^{\mathrm T}$ be a column
vector of functions $F_j(x,\yy,\uu)$, $1\le j\le N$, with complex
variables $x$, $\yy = (y_1,\ldots,y_N)^{\mathrm T}$, $\uu =
(u_1,\ldots,u_k)^{\mathrm T}$ which are analytic around $0$ and
satisfy $F_j(0,{\bf 0},{\bf 0}) = 0$ for $1\le j\le N$.  We are
interested in the analytic solution $\yy = \yy(x,\uu) =
(y_1(x,\uu),\ldots,y_N(x,\uu))^{\mathrm T}$ of the functional
equation
\begin{equation}
\yy = \FF(x,\yy,\uu)
\label{eq11}\end{equation}
with $\yy(0,{\bf 0}) = {\bf 0}$, i.e., we demand that the (unknown)
functions $y_j = y_j(x,\uu)$, $1\le j\le N$, satisfy
the system of functional equations
\begin{align*}
y_1 &= F_{1}(x,y_1,y_2,\ldots,y_N,\uu),\\
y_2 &= F_2(x,y_1,y_2,\ldots,y_N,\uu),\\
&\,\vdots \\
y_N &= F_N(x,y_1,y_2,\ldots,y_N,\uu).
\end{align*}

It is convenient to define the notion of a
dependency (di)graph $G_{\FF} = (V,E)$ for such a system of
functional equations $\yy = \FF(x,\yy,\uu)$.
The vertices $V = \{y_1,y_2,\ldots,y_N\}$
are just the unknown functions
and an ordered pair  $(y_i,y_j)$ is contained in the edge set
$E$ if and only if $F_i(x,\yy,\uu)$ really depends on $y_j$.

If the functions $F_j(x,\yy,\uu)$ have non-negative Taylor coefficients
then it is easy to see that the solutions $y_j(x,\uu)$ have the
same property. (One only has to solve the system iteratively
by setting $\yy_0(x,\uu) = 0$ and $\yy_{i+1}(x,\uu) =
\FF(x,\yy_i(x,\uu),\uu)$ for $i\ge 0$. The limit $\yy(x,\uu) =
\lim_{i\to\infty} \yy_i(x,\uu)$ is the (unique) solution of the
system above.)

Now suppose that $G(x,\yy,\uu)$ is another analytic function with
non-negative Taylor coefficients. Then $G(x,\yy(x,\uu),\uu)$ has
a power series expansion
\[
G(x,\yy(x,\uu),\uu) = \sum_{n,\mm} c_{n,\mm} x^n \uu^{\mm}
\]
with non-negative coefficients $c_{n,\mm}$. In fact, we assume that
for every $n\ge n_0$ there exists $\mm$ such that $c_{n,\mm}>0$.

Let $\XX_n$ $(n\ge n_0)$ denote an $N$-dimensional discrete random vector with
\begin{equation}\label{defXn}
{\bf Pr} [\XX_{n} = \mm] := \frac{c_{n,\mm}}{c_{n}},
\end{equation}
where
\[
c_n = \sum_{\mm} c_{n,\mm}
\]
are the coefficients of
\[
G(x,\yy(x,{\bf 1}),{\bf 1}) = \sum_{n\ge 0} c_{n} x^n.
\]
The following theorem shows that (under suitable analyticity conditions)
$\XX_n$ has a Gaussian limiting distribution.

\begin{theorem}
\label{th:syseqns} Let\/ $\FF(x,\yy,\uu) =
(a_1(x,\yy,\uu),\ldots,F_N(x,\yy,\uu))^{\mathrm T}$ be functions analytic
around $x=0$, \ $\yy = (y_1,\ldots,y_N)^{\mathrm T} = {\bf 0}$, \
$\uu = (u_1,\ldots,u_k)^{\mathrm T} = {\bf 0}$,
whose Taylor coefficients are all
non-negative, such that\/ $\FF(0,\yy,\uu) = {\bf 0}$, \
$\FF(x,{\bf 0},\uu)\neq {\bf 0}$, \ $\FF_x(x,\yy,\uu)\not
= {\bf 0}$, and such that there exists $j$ with\/
$\FF_{y_jy_j}(x,\yy,\uu)\neq {\bf 0}$. Furthermore assume that
the region of convergence of\/ $\FF$ is large enough that there
exists a non-negative solution $x=x_0$, \ $\yy=\yy_0$ of the system of
equations
\begin{align*}
\yy &= \FF(x,\yy,{\bf 1}),\\
0 &= \det(\II - \FF_\yy(x,\yy,{\bf 1})),
\end{align*}
inside it. Let
\[
\yy = \yy(x,\uu) = (y_1(x,\uu),\ldots,y_N(x,\uu))^{\mathrm T}
\]
denote the analytic solutions of the system
\begin{equation}
\yy = \FF(x,\yy,\uu)
\label{eq120}
\end{equation}
with $\yy(0,\uu)=\bf 0$ and assume that $d_{n,j}>0$ $(1\le j\le N)$
for $n\ge n_1$, where $y_j(x,{\bf 1}) = \sum_{n\ge 0}d_{n,j} x^n$.
Moreover, let $G(x,\yy,\uu)$ denote an analytic function with
non-negative Taylor coefficients such that the point $(x_0,
\yy(x_0,{\bf 1}),{\bf 1})$ is contained in the region of
convergence. Finally, let random vectors\/~$\XX_n$ $(n\ge n_0)$ be
defined by (\ref{defXn}).

If the dependency graph $G_\FF=(V,E)$ of the system
(\ref{eq120})
in the unknown functions
$y_1(x,\uu),\ldots,$ $y_N(x,\uu)$
is strongly connected then the sequence of
random vectors\/~$\XX_{n}$
admits a Gaussian limiting distribution with mean value
\[
\E\, \XX_{n} = \mmu\, n + O(1)\qquad (n\to\infty)
\]
and covariance matrix
\[
{\bf Cov}(\XX_{n},\XX_{n}) = {\bf\Sigma}\, n + O(1)\qquad
(n\to\infty).
\]
The row vector $\mmu$ is given by
\begin{equation*}
\mmu = -\frac{x_\uu({\bf 1})}{x({\bf 1})},
\end{equation*}
and the matrix $\bf\Sigma$ by
\begin{equation}\label{eqsigma}
\bf\Sigma= -\frac{x_{\uu\uu}({\bf 1})}{x({\bf 1})} + \mmu^{\mathrm T}\mmu +
{\rm diag}(\mmu),
\end{equation}
where $x=x(\uu)$ (and $\yy= \yy(\uu) = \yy(x(\uu),\uu)$) is the solution of
the (extended) system
\begin{align}
\yy &= \FF(x,\yy,\uu), \label{xu1}\\
0 &= \det(\II - \FF_\yy(x,\yy,\uu)). \label{xu2}
\end{align}
\end{theorem}

The proof of Theorem \ref{th:syseqns} is exactly the same as
that given in  \cite{Dr97}. The main observation is that
the assumptions above show that the solutions $y_j(x,\uu)$ admit
a local representation of the form
\[
y_j(x,\uu) = g_j(x,\uu) - h_j(x,\uu)\sqrt{1-\frac x{x(\uu)}},
\]
(where $\uu$ is close to ${\bf 1}$ and $x$ close to $x_0 = x({\bf 1})$).
The assumption that the dependency graph is strongly connected
ensures that the location of the singularity of all functions $y_j(x,\uu)$ is
determined by the common function $x(\uu)$. Thus, we get the same
property for $G(x,\yy(x,\uu),\uu)$:
\begin{equation}\label{eqsqrtrelation}
G(x,\yy(x,\uu),\uu) = g(x,\uu) - h(x,\uu)\sqrt{1-\frac x{x(\uu)}}
\end{equation}
It is then well known (see~\cite{MR85k:05009,Dr94}) that a
square-root singularity plus some minor conditions implies
asymptotic normality of the coefficients (in the sense introduced
above) with mean and covariance expressed in terms of derivatives of
$x(\uu)$. Note, for example, that the assumption $d_{n,j}>0$ for
$n\ge n_1$ ensures that $c_n>0$ for sufficiently large $n$ and from
this follows that $x_0=x({\bf 1})$ is the only singularity on the
radius of convergence of $G(x,\yy(x,{\bf 1}),{\bf 1})$.

\medskip

In what follows we comment on the evaluation of $\mmu$ and
$\bf\Sigma$. The problem is to extract the derivatives of $x(\uu)$.
The function~$x(\uu)$ is the solution of the system (\ref{xu1}--\ref{xu2}) and
is exactly the location of the singularity of the mapping $x\mapsto
\yy(x,\uu)$ when $\uu$ is fixed (and close to ${\bf 1}$).

Let $x(\uu)$ and $\yy(\uu) = \yy(x(\uu),\uu)$ denote the solutions
of (\ref{xu1}--\ref{xu2}). Then we have
\begin{equation}\label{zeroder}
\yy(\uu) = \FF(x(\uu),\yy(\uu),\uu).
\end{equation}
Taking derivatives with respect to $\uu$ we get
\begin{equation}\label{firstder}
\yy_\uu(\uu) = \FF_x(x(\uu),\yy(\uu),\uu) x_\uu(\uu) +
\FF_\yy(x(\uu),\yy(\uu),\uu) \yy_\uu(\uu) + \FF_\uu(x(\uu),\yy(\uu),\uu),
\end{equation}
where the three terms in~$\FF$ denote evaluations at
$(x(\uu),\yy(\uu),\uu)$ of the partial derivatives of~$\FF$,
and where $x_\uu$ and $\yy_\uu$ denote the Jacobian of $x$
resp. $\yy$ with respect to $\uu$. In particular, for $\uu =
{\bf 1}$
we have $x({\bf 1}) = x_0$ and $\yy({\bf 1}) = \yy_0$ and, of course
\[
\det(\II - \FF_\yy(x_0,\yy_0,{\bf 1})) = 0.
\]
Since $\FF_\yy$ is a non-negative matrix and the dependency graph is
strongly connected there is a unique Perron-Frobenius eigenvalue of
multiplicity 1. Here this eigenvalue equals 1. Thus, $\II - \FF_\yy$
has rank $N-1$ and has (up to scaling) a unique positive left
eigenvector $\bbb^{\mathrm T}$:
\begin{equation*}\label{bify}
\bbb^{\mathrm T}(\II - \FF_\yy(x_0,\yy_0,{\bf 1})) = {\bf 0}.
\end{equation*}
{}From (\ref{firstder}) we obtain
\begin{equation*} 
(\II - \FF_\yy(x_0,\yy_0,{\bf 1}))\yy_\uu({\bf 1}) = \FF_x(x_0,\yy_0,{\bf 1})
x_\uu({\bf 1})
+  \FF_\uu(x_0,\yy_0,{\bf 1}).
\end{equation*}
By multiplying $\bbb^{\mathrm T}$ from the left we thus get
\begin{equation}\label{bFx}
\bbb^{\mathrm T} \FF_x(x_0,\yy_0,{\bf 1}) x_\uu + \bbb^{\mathrm
T}\FF_\uu(x_0,\yy_0,{\bf 1}) = 0
\end{equation}
and consequently
\begin{equation*} 
\mmu = \frac 1{x_0}\frac {\bbb^{\mathrm T}\FF_\uu(x_0,\yy_0,{\bf
1})}{\bbb^{\mathrm T}\FF_x(x_0,\yy_0,{\bf 1})}
\end{equation*}

\medskip

The derivation of ${\bf\Sigma}$ is more involved.
We first define ${\bf b}(x,\yy,\uu)$ as the (generalized) vector
product\footnote{More precisely this is the wedge product combined with
the Hodge duality.} of the $N-1$ last columns of the matrix
${\bf I}-\FF_\yy(x,\yy,\uu)$. Observe that
\[
D(x,\yy,\uu) := \left( {\bf b}^T(x,\yy,\uu)\left({\bf I}-\FF_\yy(x,\yy,\uu)\right) \right)_1 = 
\det\left(\II -\FF_\yy(x,\yy,\uu)\right).
\]
In particular we have
\[
D(x(\uu),\yy(\uu),\uu) = 0.
\]
Then from
\begin{align}
(\II -\FF_\yy)\yy_\uu &= \FF_x x_\uu + \FF_\uu ,\nonumber \\
-D_\yy \yy_\uu &=  D_x x_\uu + D_\uu  \label{eqyu}
\end{align}
we can calculate $\yy_\uu$. (The first system has rank $N-1$, this means that
we can skip the first equation. This reduced system 
is then completed to a regular system by appending the second equation (\ref{eqyu}).)

We now set
\begin{align*}
d_1(\uu) &= d_1(x(\uu), \yy(\uu), \uu) = \bbb(x(\uu), \yy(\uu),
\uu)^{\mathrm T} \FF_x(x(\uu), \yy(\uu), \uu) \\
\dd_2(\uu) &= \dd_2(x(\uu), \yy(\uu), \uu) = \bbb(x(\uu), \yy(\uu),
\uu)^{\mathrm T} \FF_\uu(x(\uu), \yy(\uu), \uu).
\end{align*}
By differentiating equation \eqref{bFx} we get
\begin{equation}\label{eqxuu}
x_{\uu\uu}(\uu) = -\frac{(d_{1x} x_\uu + d_{1\yy} \yy_\uu +
d_{1\uu}) x_\uu + (\dd_{2x} x_\uu + \dd_{2\yy} \yy_\uu +
\dd_{2\uu})} {d_1},
\end{equation}
where $d_{1x},d_{1\yy},d_{1\uu},\dd_{2x},\dd_{2\yy},\dd_{2\uu}$
denote the respective partial derivatives and where we omitted the
dependence on $\uu$. With the knowledge of $x_0,\yy_0$ and
$\yy_\uu({\bf 1})$ we can now evaluate $x_{\uu\uu}$ at $\uu = {\bf
1}$ and we finally calculate $\bf\Sigma$ from \eqref{eqsigma}.

\section{Proof of Lemma~\ref{determinant}}\label{sec-determinant}

In this appendix we will prove Lemma~\ref{determinant} saying that the determinant 
$\det\left({\bf I} - {\bf F}_\aaa(x,\aaa,1)\right)$ is given by
\[
\det\left({\bf I} - {\bf F}_\aaa(x,\aaa,1)\right) = 1 - x e^{a_0+a_1+ \cdots+ a_L}.
\]

We first observe that the sum of all rows of ${\bf I} - {\bf F}_\aaa(x,\aaa,1)$ 
equals
\[
\left( 1 - x e^{a_0+a_1+ \cdots+ a_L}, 1 - x e^{a_0+a_1+ \cdots+ a_L}, \ldots, 
1 - x e^{a_0+a_1+ \cdots+ a_L} \right),
\]
compare with (\ref{eqmatrix1}). Hence, we get
\[
\det\left({\bf I} - {\bf F}_\aaa(x,\aaa,1)\right) = (1-xe^{a_0+a_1+\ldots+a_L}) \det {\bf M}(x,\aaa),
\]
where ${\bf M}(x,\aaa)$ denotes the matrix ${\bf I} - {\bf F}_\aaa$ where we replace the
first row by $(1,1,\ldots,1)$.
Thus, it remains  to prove that $\det {\bf M}(x,\aaa) = 1$. 

For this purpose we have to be more explicit with the partition $\mathcal A=\{a_0,a_1,\dots,a_L\}$.
More precisely we construct $\mathcal A$ recursively from level to level.
This procedure is similar to that of Proposition~\ref{Pro3} but not the same. In order to make our
arguments more transparent we restrict ourselves to 4~steps. Note that this procedure
also provides a recursive description of the polynomials~$P_j(\aaa,1)$.

One starts with $\mathcal A_0=\{d_0, d_1\}$, where $d_0 = a_0$ and $d_1 = p\setminus a_0$. This 
means that $d_0$ collects all trees where the root out-degree is not contained in~$D$ and $d_1$ those
where it is contained in~$D$. For example, if $D = \{2\}$ then the generating functions
of this (trivial) partition are given by $d_1(x,1) = x p(x)^2/2$ and by $d_0(x,1) = p(x) - d_1(x,1)
= p(x) -x p(x)^2/2$.

Then we partition $d_1$ according to structure of the subtrees of the root, where
we distinguish between the previous classes $d_0$ and $d_1$. We 
get $\mathcal A_1=\{c_0, c_1,\ldots,c_m\}$, where $c_0 = d_0$ and $c_1 \oplus \ldots \oplus c_m = d_1$.
In particular, if  $D = \{2\}$ then $m=3$, the class $c_1$ collects all trees 
with root out-degree $2$ where both subtrees
of the root are in class $a_0=d_0$, $c_2$ collects all trees with with root out-degree $2$ where
one subtree of the root is in class $a_0 = d_0$ and the other one in class $d_1$, and $c_3$ collects 
those trees
where both subtrees of the root are in class $d_1$. The corresponding generating functions
are given by $c_1(x,1) = x d_0(x,1)^2/2$, by  $c_2(x,1) = x d_0(x,1)d_1(x,1)$, and by
$c_3(x,1) = x d_1(x,1)^2/2$. Of course, we also have $c_0(x,1) = d_0(x,1)$ and
$c_1(x,1) + c_2(x,1) + c_3(x,1) = d_1(x,1)$.

In the same fashion we proceed further. 
We partition $c_s$ ($1\le s\le m$) according to the structure
of the subtrees of the root (that are now taken from $\{c_1,\ldots,c_m\}$) 
and denote them by $\mathcal A_2=\{b_0, b_1,\ldots,b_\ell\}$. 
Further we define sets $C_s$ by $c_s = \bigoplus_{r\in C_s} b_r$.
If $D = \{2\}$ then $b_0 = c_0$, $b_1 = c_1$, $c_2$ is divided into three parts, and
$c_3$ is divided into 6 parts: $C_1 = \{1\}$, $C_2 = \{2,3,4\}$, $C_3 = \{5,6,7,8,9,10\}$.\footnote
{By the way this leads to the partition 
that is used in the proof of Theorem~\ref{Th1} resp.\ of Proposition~\ref{Pro1}.}

Finally, we partition 
$b_j$ ($j\ge 1$) according according to the structure
of the subtrees of the root that are taken from the $b_i$ and denote 
them by $\mathcal A =\{a_0, a_1,\ldots,a_L\}$. 
As in the previous step we define sets $B_r$ by
$b_r = \bigoplus_{j\in B_r} a_j$.
In general we have to iterate this procedure until a certain level and
get almost the same partition as in the proof of Proposition~\ref{Pro1}. The only difference is that at the lowest level we only distinguish between nodes with degree in $D$ and degree not in $D$. However this is no real restriction as we can extend the partition above with an additional level and we will have a well-defined number of additional occurrences for each class. We again obtain  a partition which fits Proposition~\ref{Pro1}.

We recall that this recursive procedure directly provides
a recursive description of the system of functional equations. In particular
we have
\[
a_j(x,1) = x\,P_j(a_0(x,1),a_1(x,1),\ldots,a_L(x,1),1),
\]
where $P_j(\cdot,1)$ can be actually written as a polynomial in $b_0,b_1,\ldots,b_\ell$. 

Next
\[
b_r(x,1) = x \,Q_r(b_0(x,1),b_1(x,1),\ldots,b_\ell(x,1),1),
\]
where $Q_r(\cdot,1)$ can be written as a polynomial in $c_0,c_1,\ldots, c_m$. Further,
\[
Q_r = \sum_{j\in B_r} P_j.
\]
In other words, the sum $\sum_{j\in B_r} P_j$ can be written as polynomial in $c_r$.

Finally, 
\[
c_s(x,1) = x \,R_s(c_0(x,1),c_1(x,1),\ldots,c_m(x,1)),
\]
where $R_s(\cdot,1)$ can be written as a polynomial in $d_0 = a_0$ and $d_1 = a_1+ \cdots +a_L$ and 
we have
\[
R_s = \sum_{r\in C_s} Q_r.
\]

Let ${\bf G}(x,\aaa)$ denote the $L\times L$-submatrix of $\FF_\aaa$ where we omit the first row and column.
Then ${\bf G}(x,\aaa)$ has the following structure:
\[
{\bf G}(x,\aaa) = \left( \begin{array}{ccc} G_{11} & \cdots & G_{1m} \\ \vdots & & \vdots \\ 
 G_{m1} & \cdots & G_{mm} \end{array} \right),
\]
where
\[
G_{s's''} = \left( \begin{array}{c} B_{r'r''}  \end{array} \right)_{r'\in C_{s'}, r''\in C_{s''}}
\]
and 
\[
B_{r'r''} =  \left( \begin{array}{c} x P_{i,a_j} \end{array} \right)_{i\in B_{r'},j\in B_{r''}}.
\]

The condition that $P_i$ can be written as a polynomial in $b_j$ implies that 
$P_{i,a_{j_1}} = P_{i,a_{j_2}}$ for all $j_1,j_2\in B_{r''}$, that is,
each row of $B_{r'r''}$ is either zero or all entries are the same.

Further, if we fix  $r'$ and sum over all rows $i \in B_{r'}$ then
we get
\[
\sum_{i\in B_{r'}} x P_{i,a_j} = x Q_{r',a_j}.
\]
Since $Q_{r'}$ can be written as a polynomial in $c_s$ ($0\le s\le m$) we have
$Q_{r',a_{j_1}} = Q_{r',a_{j_2}}$ for all $j_1,j_2 \in \bar C_{s''}$, where we set $\bar C_s = \bigcup\limits_{r\in C_s} B_r$.

Similarly if we fix $s'$ and sum over all rows $i \in \bar C_{s'}$ then 
we get
\[
\sum_{i\in \bar C_{s'}} x P_{i,a_j} = x R_{s',a_j}.
\]
Since $R_{s'}$ can be written as a polynomial in $d_0=a_0$ and $d_1 = a_1 + \cdots + a_L$
we have $R_{s',a_{j_1}} = R_{s',a_{j_2}}$ for all $1\le j_1,j_2\le L$.

Now we will calculate the determinant of the matrix
\begin{align*}
{\bf M}(x,\aaa) &=  \left( \begin{array}{cccc} 1 & 1 \cdots  & \cdots &  \cdots 1 \\
                                              0 &  {\bf I} & \cdots & {\bf 0} \\
                                              \vdots &  \vdots & \ddots & \vdots \\
                                             0&  {\bf 0} & \cdots & {\bf I} 
                                             \end{array} \right) -                                              
  \left( \begin{array}{cccc}  0 & {\bf 0} & \cdots & {\bf 0} \\
  \times & G_{11} & \cdots & G_{1m} \\ 
  \vdots &\vdots & & \vdots \\ 
  \times & G_{m1} & \cdots & G_{mm}  \end{array} \right) \\ 
& =  \left( \begin{array}{cccc} 1 & 1 \cdots & \cdots &  \cdots 1 \\                                         
  \times &  {\bf I}- G_{11} & \cdots & -G_{1m} \\ 
  \vdots &\vdots & & \vdots \\ 
  \times & -G_{m1} & \cdots & {\bf I}-G_{mm}  \end{array} \right).
\end{align*}
(By $\times$ we denote an entry we do not care.)
We now perform the following row operations. For every $s' = 1,\ldots, m$
we substitute the first row of 
\[
\left( \begin{array}{cccccc}                                        
  \times &  - G_{s'1} & \cdots &  {\bf I}- G_{s's'} & \cdots  -G_{s'm} 
 \end{array} \right)   
\]
by the sum of the corresponding rows $i\in \bar C_{s'}$. Since $R_{s',a_{j_1}} = R_{s',a_{j_2}}$ for all $1\le j_1,j_2\le L$
this sum of the rows has the form
\[
\left( \begin{array}{cccccc}                                        
  \times &  - x R_{s',a} \cdots  -x R_{s',a} & \cdots & 1- x R_{s',a} \cdots  1-x R_{s',a} & 
  \cdots & - x R_{s',a} \cdots  -x R_{s',a} 
 \end{array} \right) 
\]
We now add  the very  first row (that equals $(1,1,\ldots,1)$) $xR_{s',a}$ times to this row and obtain
\[
{\bf w}_{s'} = \left( \begin{array}{ccccccccccc}                                        
  \times &|& 0\cdots  0 &|& \cdots &|& 1 \cdots  1 &|&  \cdots &|& 0 \cdots  0 
 \end{array} \right)
\]
Next we fix  $s'$ and $r'$ such that $r'\in C_{s'}$ and substitute the first row of
\[
\left( \begin{array}{cccccc}                                        
  \times &  (- B_{r'j})_{j\in C_1}  & \cdots &  ({\bf I}\cdot\delta_{r'j}- B_{r'j})_{j\in C_{s'}} & \cdots  
  (- B_{r'j})_{j\in C_m}
 \end{array} \right)   
\]
by the sum of the rows $i\in B_{r'}$. Since for every $s''$ it holds that $Q_{r',a_{j_1}} = Q_{r',a_{j_2}}$ for all $j_1,j_2 \in \bar C_{s''}$
this sum has the following form
\[
\left( \begin{array}{cccccc}                                        
  \times &  (- x Q_{r',a_j})_{j\in \bar C_1}  & \cdots &  (\bar\delta_{r'j} - x Q_{r',a_j}  )_{j\in \bar C_{s'}} & \cdots  
  (- x Q_{r',a_j})_{j\in \bar C_m} 
 \end{array} \right),
\]
where $\bar\delta_{r'j}=1$ if and only if $j\in B_{r'}$ and $=0$ otherwise.
This means, for every $s''\ne s'$ the entries $(- x Q_{r',a_j})_{j\in \bar C_{s''}}$ are either all equal or 
if $s'' = s'$ then we have  to add 1 at proper positions. For every $s''$ we now add row ${\bf w}_{s''}$
$x Q_{r',a_j}$ times. If $s''\ne s'$ then we get a zero block $(0,\ldots, 0)$. If $s'' = s'$ we get a block of 
the form 
\[
\left( \begin{array}{ccccc}                                        
   0\cdots 0 & \cdots & 1\cdots 1 & \cdots & 0\cdots 0
 \end{array} \right).
\]
This means that this row is replaced by
\[
{\bf w}_{s',r'} = \left( \begin{array}{ccccccccccccccc}                                        
 \times &|& 0 \cdots 0 &|& \cdots &|& 0 \cdots 0 &|&  0\cdots 0 \  \cdots \ 1\cdots 1 \ \cdots \  0\cdots 0 &|& 0 \cdots 0 &|& \cdots &|& 0 \cdots 0 
 \end{array} \right).
\]
With help of these rows we can eliminate all further entries of ${\bf M}(x,\aaa)$ that come
from ${\bf G}(x,\aaa)$. (Here we use the fact that each row of $B_{r'r''}$ 
is either zero or all entries are the same.) This means that we finally end up with a matrix of
the form
\[
{\bf H} = \left( \begin{array}{cccc} 1 & 1 \cdots & \cdots &  \cdots 1 \\                                         
  \times &  H_{11} & \cdots & H_{1m} \\ 
  \vdots &\vdots & & \vdots \\ 
  \times & H_{m1} & \cdots & H_{mm}  \end{array} \right),
\]
where $H_{s's''} = {\bf 0}$ for $s' \ne s''$ and 
$H_{s's'}$ is of the form
\[
H_{s's'} = \left( \begin{array}{ccccc} J & K  & K & \cdots &   K \\                                         
  {\bf 0} &  J & {\bf 0}& \cdots & {\bf 0} \\ 
    \vdots & & \ddots & & \vdots \\ 
 \vdots & & & \ddots & \vdots\\
  {\bf 0} & {\bf 0} &  {\bf 0} &\hdots & J \end{array} \right).
\]
with
\[
J= \left( \begin{array}{ccccc} 1 & 1  &1 & \cdots  &  1 \\                                         
  0 &  1 & 0 & \cdots & 0 \\ 
  \vdots & &  \ddots & & \vdots \\
  \vdots & & & \ddots & \vdots \\ 
  0 & 0 &  0 & \hdots &  1  \end{array} \right)
\quad\mbox{and} \quad
 K= \left( \begin{array}{ccccc} 1 & 1  & 1 & \cdots &   1 \\                                         
  0 &  0 &  0& \hdots & 0 \\ 
  \vdots & \vdots & \vdots & & \vdots \\ 
  \vdots & \vdots & \vdots & &\vdots\\
  0 & 0 & 0& \hdots & 0  \end{array} \right).
\]
It is now an easy task to transform the matrix $(H_{s's''})_{1\le s',s''\le m}$ 
(with help of row transforms) to the identity matrix. Furthermore we can
transform the very first row $(1,1,\ldots,1)$ of ${\bf H}$ to $(1,0,\ldots,0)$ and end up with a matrix of the form
\[
\left( \begin{array}{cccc} 1 & 0  & \cdots &   0 \\                                         
  \times &  1 &  & 0 \\ 
  \vdots & & \ddots & \vdots \\ 
  \times & 0 &  & 1  \end{array} \right).
\]
Obviously, this matrix has determinant $1$. Since the above row transforms do not change the value of the
determinant we, thus, obtain $\det {\bf M}(x,\aaa) = 1$.

\end{appendix}

\medskip\noindent
{\bf Acknowledgement.} The authors want to thank Philippe Flajolet for
several discussions on the topic of the paper and for many useful hints.

\bibliography{treepatterns}
\bibliographystyle{alpha}

\end{document}